\documentclass[prd,twocolumn,showpacs,floatfix,amsmath,nofootinbib,amssymb,floatfix]{revtex4}
\usepackage{graphicx,color,dcolumn,booktabs,bm}
\usepackage{longtable,lscape}
\usepackage{txfonts}
\usepackage{overpic}
\usepackage{amssymb}
\usepackage{array}
\usepackage{indentfirst}
\usepackage{feynmf}   
\usepackage{slashed}  
\usepackage{cases}
\usepackage{color}
\usepackage{multirow}
\usepackage{epstopdf}
\usepackage{longtable}
\usepackage{graphicx,color,dcolumn,booktabs,bm,float}
\graphicspath{{Figures/}{logo/}}
\usepackage[colorlinks,
            citecolor=blue,
            anchorcolor=red,
            menucolor=red,
            linkcolor=red,
            filecolor=red,
            runcolor=red,
            urlcolor=blue,
            frenchlinks=red]{hyperref}

\graphicspath{{Figures/}} %

\allowdisplaybreaks

\begin{document}

\title{Exploring two-body strong decay properties for possible single charm molecular pentaquarks with strangeness $|S|=1,2$}

\author{Xiao-Mei Tang$^{1}$}
\author{Jin-Yu Huo$^{1}$}
\author{Qi Huang$^{2}$}\email{06289@njnu.edu.cn}
\author{Rui Chen$^{1,3}$}\email{chenrui@hunnu.edu.cn}

\affiliation{
$^1$Key Laboratory of Low-Dimensional Quantum Structures and Quantum Control of Ministry of Education, Department of Physics and Synergetic Innovation Center for Quantum Effects and Applications, Hunan Normal University, Changsha 410081, China\\
$^2$Department of Physics and Technology, Nanjing Normal University, Nanjing 210023, China\\
$^3$Hunan Research Center of the Basic Discipline for Quantum Effects and Quantum Technologies, Hunan Normal University, Changsha 410081, China}
\date{\today}

\begin{abstract}

The exploration of exotic hadrons provides a crucial testing ground for quantum chromodynamics in its non-perturbative regime. In this work, we perform a systematic study of the two-body strong decay properties of single-charm molecular pentaquarks in the $Y_c\bar{K}^{(*)}$ systems, where $Y_c = \Lambda_c$, $\Sigma_c$, $\Xi_c$, and $\Xi_c'$. Employing an effective Lagrangian approach combined with hadronic molecular wave functions derived from the one-boson-exchange model, we compute the decay widths and branching ratios for a series of predicted states with strangeness $|S| = 1$ and $|S| = 2$. Our calculations reveal distinctive decay patterns that serve as fingerprints for molecular identification. The total decay widths vary dramatically, from less than 1 MeV for the narrow $\Sigma_c\bar{K}$ $(I(J^P)=1/2(1/2^-))$ state to several tens of MeV for broader coupled-channel molecules like $\Lambda_c\bar{K}^*/\Sigma_c\bar{K}^*$. A key finding is the stability of the predicted branching ratios against variations in the binding energy. The decay dynamics are dominated by light meson (particularly pion) exchange, leading to a strong preference for final states containing a charmed baryon and a strange meson. Furthermore, coupled-channel effects and isospin-related interference play essential roles in both the formation and decay mechanisms of specific candidates. The results provide concrete, testable predictions for future experimental searches at facilities such as LHCb and Belle II. 

\end{abstract}


\maketitle

\section{introduction}

Since M. Gell-Mann and G. Zweig first proposed exotic states beyond the conventional mesons and baryons in 1964 \cite{Gell-Mann:1964ewy,Zweig:1964ruk,Zweig:1964jf}, both theorists and experimentalists have devoted sustained effort to their discovery and identification. The past two decades, in particular, have witnessed a proliferation of new hadron states reported by experiments (see comprehensive reviews \cite{Chen:2016qju, Guo:2017jvc, Chen:2022asf, Liu:2019zoy, Chen:2016spr, Liu:2013waa, Hosaka:2016pey, Meng:2022ozq}). These observations have prompted a wide array of theoretical interpretations for their internal structure, ranging from conventional quark-model states to exotic configurations such as compact multiquarks, hybrids, glueballs, and hadronic molecules.

A prime example is the spectroscopy of charmed-strange baryons. In 2017, the LHCb Collaboration observed five narrow $\Omega_c^0$ resonances—$\Omega_c^0(3000)$, $\Omega_c^0(3050)$, $\Omega_c^0(3066)$, $\Omega_c^0(3090)$, and $\Omega_c^0(3119)$—in the $\Xi_c^+K^-$ mass spectrum \cite{LHCb:2017uwr}. These findings were soon corroborated by the Belle Collaboration, which confirmed four of the five states using their full data set \cite{Belle:2017ext}. Further refinement came in 2023, when LHCb not only confirmed all five original states but also reported two new resonances, $\Omega_c(3185)$ and $\Omega_c(3327)$ \cite{LHCb:2023sxp}.

Theoretical explanations for these states are diverse. While many studies interpret them as conventional excited charm baryons within the quark model \cite{
Weng:2024roa,Patel:2023wbs,Ortiz-Pacheco:2020hmj,
Yang:2023fsc,Cheng:2015naa,Chen:2017sci,Karliner:2017kfm,Wang:2017hej,
Wang:2017vnc,Padmanath:2017lng,Cheng:2017ove,Wang:2017zjw,
Zhao:2017fov,Chen:2017gnu,Yu:2022ymb,Luo:2023sra,
Peng:2024pyl,Agaev:2017lip,Faustov:2020gun,Jia:2020vek,
Polyakov:2022eub,Yu:2023bxn,Wang:2023wii,Oudichhya:2023awb,
Ortiz-Pacheco:2023kjn,Pan:2023hwt,Li:2024zze,Ortiz-Pacheco:2024qcf,Zhong:2025oti,Agaev:2017jyt}, alternative exotic interpretations also exist \cite{Feng:2023ixl,Luo:2023sne,Zhang:2025gar,Huang:2017dwn,Kim:2017jpx,Liu:2017frj,Kim:2017khv,Debastiani:2017ewu,Wang:2017smo,Chen:2017xat,Nieves:2017jjx,Huang:2018wgr,Debastiani:2018adr,Wang:2018alb,Xu:2019kkt,Zhu:2022fyb,Ozdem:2023okg,Wang:2023eng,Ikeno:2023uzz,Xin:2023gkf}. Notably, the $\Omega_c^0(3050)$ and $\Omega_c^0(3090)$ have been proposed as baryon-meson molecular states. Furthermore, analyses in Refs. \cite{An:2017lwg,Yang:2017rpg,Montana:2017kjw,Yan:2023tvl} suggest that three of the states may be compact pentaquarks $(sscq\bar{q})$ with masses near $\Omega_c^0(3066)$, $\Omega_c^0(3090)$, and $\Omega_c^0(3119)$. In our previous work \cite{Chen:2017xat}, we identified a loosely bound $\Omega_c$-like molecular state, predominantly composed of $\Xi_c^*\bar{K}$ with $I(J^P)=0(3/2^-)$, within the one-boson-exchange (OBE) model—a state potentially related to $\Omega_c^0(3119)$.

Distinguishing between conventional and exotic interpretations for such states is a crucial and pressing challenge in hadron spectroscopy. Decay properties can offer a powerful diagnostic tool in this regard, as they are highly sensitive to the underlying wave functions. For instance, the branching fractions of excited $\Omega_c^0$ baryons into channels such as $\Xi_c\pi$ or $\Omega_c\eta$ are expected to differ significantly between molecular and compact pentaquark scenarios \cite{Cheng:2006dk,Wang:2017hej,Chen:2007xf,Zhong:2007gp}.

In a recent study, we employed the one-boson-exchange (OBE) model to investigate $Y_c\bar{K}^{(*)}$ $(Y_c=\Lambda_c, \Sigma_c)$ interactions \cite{Chen:2023qlx}. Incorporating both $S$-$D$ wave mixing and coupled-channel effects, we predicted several charmed molecular pentaquarks with strangeness $|S|=1$, including $\Sigma_c\bar{K}$ with $I(J^P)=1/2(1/2^-)$, $\Sigma_c\bar{K}^*$ with $1/2(1/2^-, 3/2^-)$ and $3/2(3/2^-)$, and the coupled $\Lambda_c\bar{K}^*/\Sigma_c\bar K^*$ molecule with $1/2(1/2^-)$. A key complication is that these predicted molecular states can share the same masses and quantum numbers with known charm-strange baryons $\Xi_c^{(')}$, making their unambiguous experimental identification based solely on mass spectroscopy challenging.

To address this challenge, we undertake in this work a systematic study of the two-body strong decay properties for possible single-charm molecular pentaquarks $Y_c\bar{K}^{(*)}$, composed of a charmed baryon $Y_c=(\Lambda_c, \Sigma_c, \Xi_c, \Xi_c^{\prime})$ and an anti-strange meson $(\bar{K}, \bar{K}^*)$. We focus on decay modes mediated by $S$-wave interactions, which can typically dominate over those involving higher partial waves ($P$-wave, $D$-wave, etc.), the latter will be addressed in future work.

Besides decay analysis, we also establish the mass spectrum of possible $\Xi_c^{(\prime)}\bar{K}^{(*)}$ molecules. We first employ the OBE model with $S$-$D$ wave mixing and coupled-channel effects to study $\Xi_c^{(\prime)}\bar{K}^{(*)}$ interactions; the detailed OBE potentials and numerical results for the mass spectrum are provided in Appendix \ref{app}. Our calculations predict several double-strange charm molecular candidates, including a coupled $\Xi_c^{\prime}\bar{K}/\Xi_c\bar{K}^*/\Xi_c^{\prime}\bar{K}^*$ state with $I(J^P)=0(1/2^-)$, coupled $\Xi_c\bar{K}^*/\Xi_c^{\prime}\bar{K}^*$ states with $0(1/2^-, 3/2^-)$, and single-channel $\Xi_c^{\prime}\bar{K}^*$ molecules with $0(1/2^-, 3/2^-)$ and $1(1/2^-)$.

With these identified molecular candidates, we next compute their two-body strong decay widths using an effective Lagrangian approach. The resulting decay patterns can provide crucial information for elucidating the internal structure of these baryon-like states, offering a means to distinguish novel hadronic configurations from conventional excitations. This study cannot only enhance our understanding of hadron spectra but also provide concrete predictions to guide future experimental analyses.

This paper is organized as follows. In Sec.~\ref{sec2}, we present the formalism for calculating the two-body strong decays of the $Y_c\bar{K}^{(*)}$ molecular pentaquarks. The numerical results for the decay properties are detailed in Sec.~\ref{sec3}. Finally, a summary is given in Sec.~\ref{sec4}.

\section{Theoretical Framework for Decay Widths}\label{sec2}

The two-body strong decays of a hadronic molecule can be effectively described at the hadronic level by the exchange of mesons or baryons between its constituent hadrons. For a molecular state that is a mixture of several coupled channels, the transition matrix element from the initial molecular state $|i\rangle$ to a final two-hadron state $|f_1 f_2\rangle$ is related to the scattering amplitudes of its constituent channels. Specifically,
\begin{eqnarray}
    \langle f_1 f_2 | \hat{T} | i\rangle &=& \sum_n \langle f_1 f_2 | \hat{T} | A_n B_n \rangle \langle A_n B_n | i\rangle\nonumber\\
    &=& \sum_n \int \frac{d^3 \bm{k}^\prime d^3 \bm{k}}{(2\pi)^3} \delta(\bm{p}-\bm{k}^\prime) \langle \bm{k}^\prime | \hat{T} | \bm{k} \rangle \psi_{i,A_n B_n}(\bm{k})\nonumber\\
    &=& \sum_n \int \frac{d^3 \bm{k}}{(2\pi)^3} \sqrt{\frac{2E_i}{4E_{A_n}E_{B_n}}}\mathcal{M}\left(A_n B_n \to f_1 f_2\right)\nonumber\\
    &&\times\psi_{i,A_n B_n}(\bm{k}).
\end{eqnarray}
Here, $E_i$ is the energy of particle $i$, $\hat{T}$ is the transition operator of $A_n + B_n \to f_1 + f_2$ process through the exchange of an intermediate hadron, $\psi_{A_nB_n}(\bm{k})$ is the wave function of the $n$-th constituent channel $|A_n B_n\rangle$ in momentum space with normalization condition  $\int d^3\bm{k}|\psi(\bm{k})|^2 = (2\pi)^3$. To account for the composite structure of the hadrons and to regularize the high-momentum behavior, we introduce a Gaussian-type form factor into the scattering amplitude at each interaction vertex~\cite{Yue:2024paz}:
\begin{eqnarray}
\mathcal{F}(m, q, \Lambda) = \frac{\Lambda^{4}}{(m^{2} - q^{2})^{2} + \Lambda^{4}},
\end{eqnarray}
where $m$ and $q$ are the mass and four-momentum of the exchanged hadron, respectively. Following previous studies~\cite{Yue:2024paz}, we adopt a cutoff value of $\Lambda = 1.00$ GeV.

Usually, the wave function in momentum representation can be related to the solution of Schr\"{o}dinger equation, which is obtained from a spectrum calculation, as
\begin{eqnarray}
    \psi(\bm{k}) = \int d^3\bm{r} e^{-i \bm{k}\cdot\bm{r}}\psi(\bm{r}).
\end{eqnarray}
For example, if $\psi(\bm{r})$ is expanded through Gaussian basis
\begin{eqnarray}
    \psi(\bm{r}) = \sum_n c_n \sqrt{\frac{2^{l+2}(2\nu_n)^{l+3/2}}{\sqrt{\pi}(2l+1)!!}} e^{-\nu_n r^2} \mathcal{Y}_{l}(\bm{r}),
\end{eqnarray}
with $\mathcal{Y}_{l}(\bm{r})$ being the solid harmonic function. then $\psi(\bm{k})$ will be inferred as
\begin{eqnarray}
    \psi(\bm{k})=\sum_n c_n \sqrt{(2\pi)^3} (-i)^l \sqrt{\frac{2^{l+2}(\frac{1}{2\nu_n})^{l+3/2}}{\sqrt{\pi}(2l+1)!!}} e^{-\frac{k^2}{4\nu_n}} \mathcal{Y}_{l}(\bm{k}).
\end{eqnarray}

To obtain the $S-$wave decay width, a partial wave projection on the decay amplitude should be done
\begin{eqnarray}
\mathcal{M}^{S L}\left(|\bm{p}|\right)&= & \sum_{M_{J_B}, M_{J_C}, M_S, M_L}\left\langle L M_L S M_S | J_A M_{J_A}\right\rangle\nonumber\\
&&\times\left\langle J_B M_{J_B} J_C M_{J_C} | S M_S\right\rangle \nonumber\\
&& \times \int d \hat{\bm{p}} Y_{L M_L}^*(\hat{\bm{p}}) \mathcal{M}^{M_{J_A} M_{J_B} M_{J_C}}(\bm{p}), 
\end{eqnarray}
where $\bm{p}$ is the three momenta of final particles in center of mass frame, and the following formula that represents a rotation on the spherical harmonic function is adopted
\begin{eqnarray}
    Y_{lm}(\theta^\prime, \phi^\prime) = \sum_{m^\prime}Y_{l m^\prime}(\theta,\phi)D_{m^\prime m}^{l}(\alpha,\beta,\gamma).
\end{eqnarray}
Here, $D_{m^\prime m}^{l}(\alpha,\beta,\gamma)$ is the Wigner $D$-function, with $\alpha$, $\beta$, and $\gamma$ being the relative Euler angles between $(\theta^\prime, \phi^\prime)$ and $(\theta,\phi)$. Finally, the partial decay width is calculated as
\begin{eqnarray}
\Gamma^{SL} &=& \frac{1}{2J+1}\frac{|\bm{p}|}{32\pi^2 m_{MS}^2}|\mathcal{M}^{SL}|^2.
\end{eqnarray}

To compute the scattering amplitudes $\mathcal{M}\left(A_n B_n \to f_1 f_2\right)$, we employ effective Lagrangians consistent with $SU(3)$ flavor symmetry~\cite{Wang:2022oof}:
\begin{eqnarray}
\mathcal{L}_{PPV} &=& i\sqrt{2}g_{PPV}(P\partial^{\mu}{P}-\partial^{\mu}{P}{P}){V}_{\mu},\label{lag0}\\
\mathcal{L}_{VVP} &=&\frac{g_{VVP}}{m_V} \epsilon_{\mu\nu\alpha\beta}
    \partial^{\mu}{V}^{\nu}\partial^{\alpha}{V}^{\beta}{P},\\
\mathcal{L}_{VVV} &=& ig_{VVV}
    \langle{V}^{\mu}[{V}^{\nu},\partial_{\mu}V_\nu]\rangle,\\
\mathcal{L}_{BBP} &=& \frac{g_{BBP}}{m_P}\bar{{B}}\gamma^\mu\gamma^5\partial_\mu {P}{B},\label{lag1}\\
\mathcal{L}_{BBV} &=& -g_{BBV}\bar{{B}}(\gamma^{\mu}
              -\frac{\kappa}{2m_{B}}\sigma^{\mu\nu}\partial_{\nu}){V}_{\mu}{B},\label{lag2}\\
\mathcal{L}_{BDP} &=&- \frac{g_{BDP}}{m_{{P}}}
             (\bar{{B}}\partial^{\mu}{P}{{D}}_{\mu}+\bar{D}_{\mu}\partial^{\mu}{P}B),\\\nonumber\\
\mathcal{L}_{BDV} &=& i\frac{g_{BDV}}{m_{{V}}}
        [\bar{{B}}\gamma_{\mu}\gamma^5{{D}}_{\nu}(\partial^{\mu}{V}^{\nu}-\partial^{\nu}{V}^{\mu}) \nonumber\\ 
        &&+\bar{{D}}_{\mu}\gamma_{\nu}\gamma^5{B}(\partial^{\mu}{V}^{\nu}-\partial^{\nu}{V}^{\mu})],\label{lag3}
\end{eqnarray}
where ${P}$, ${V}$, ${B}$, and ${D}$ denote the pseudoscalar meson, vector meson, octet baryon, and decuplet baryon fields, respectively. In Eqs. (\ref{lag0})-(\ref{lag3}), the masses $m_P$ and $m_V$ refer to the specific pseudoscalar and vector mesons involved in the vertex, rather than the universal $m_\pi$ or $m_\rho$ of the exact $SU(3)$ limit. This serves as a phenomenological correction for the significantly broken $SU(4)$ flavor symmetry in the charm sector \cite{Shen:2019evi, Yalikun:2021dpk, Wang:2022oof, Yue:2024paz}. The coupling constants in the above Lagrangians are determined by relating them to well-established interactions such as $\rho\pi\pi$, $NN\pi$, $NN\rho(\omega)$, $N\Delta\pi$, and $N\Delta\rho$. The relevant relations are summarized in Appendix \ref{app}. 

With these preparations, we proceed to analyze the two-body strong decays of the predicted $\Lambda_c\bar{K}^{(*)}$ and $\Sigma_c\bar{K}^{(*)}$ molecular pentaquarks from our previous study \cite{Chen:2023qlx}. The specific candidates under investigation are:
\begin{itemize}
    \item  $\Sigma_c\bar{K}$ molecule with $I(J^P)=1/2(1/2^-)$.
    \item $\Sigma_c\bar{K}^*$ molecules with $I(J^P)=1/2(1/2^-,3/2^-)$ and $3/2(3/2^-)$.
    \item   Coupled $\Lambda_c\bar{K}^*/\Sigma_c\bar{K}^*$ molecule with $I(J^P)=1/2(1/2^-)$.
\end{itemize}

For the charmed molecular candidates with double strangeness ($|S|=2$), a prior step is required to establish their existence and properties. We first employ the OBE model to investigate the mass spectrum of $\Xi_c^{(\prime)}\bar{K}^{(*)}$ systems. The detailed derivation of the OBE effective potentials and the numerical results for these systems are provided in Appendix \ref{app}. Our analysis identifies several promising double-strange molecular candidates using the same parameter set: 
\begin{itemize}
    \item Coupled $\Xi_c^{\prime}\bar{K}/\Xi_c\bar{K}^*/\Xi_c^{\prime}\bar{K}^*$ molecule with $I(J^P)=0(1/2^-)$.
    \item $\Xi_c^{\prime}\bar{K}^*$ molecules with  $I(J^P)=0(1/2^-, 3/2^-)$ and $1(1/2^-)$.
    \item Coupled $\Xi_c\bar{K}^*/\Xi_c^{\prime}\bar{K}^*$ molecules with $I(J^P)=0(1/2^-, 3/2^-)$.
\end{itemize}

\begin{figure*}[!htbp]
     \centering
        \includegraphics[width=0.9\linewidth]{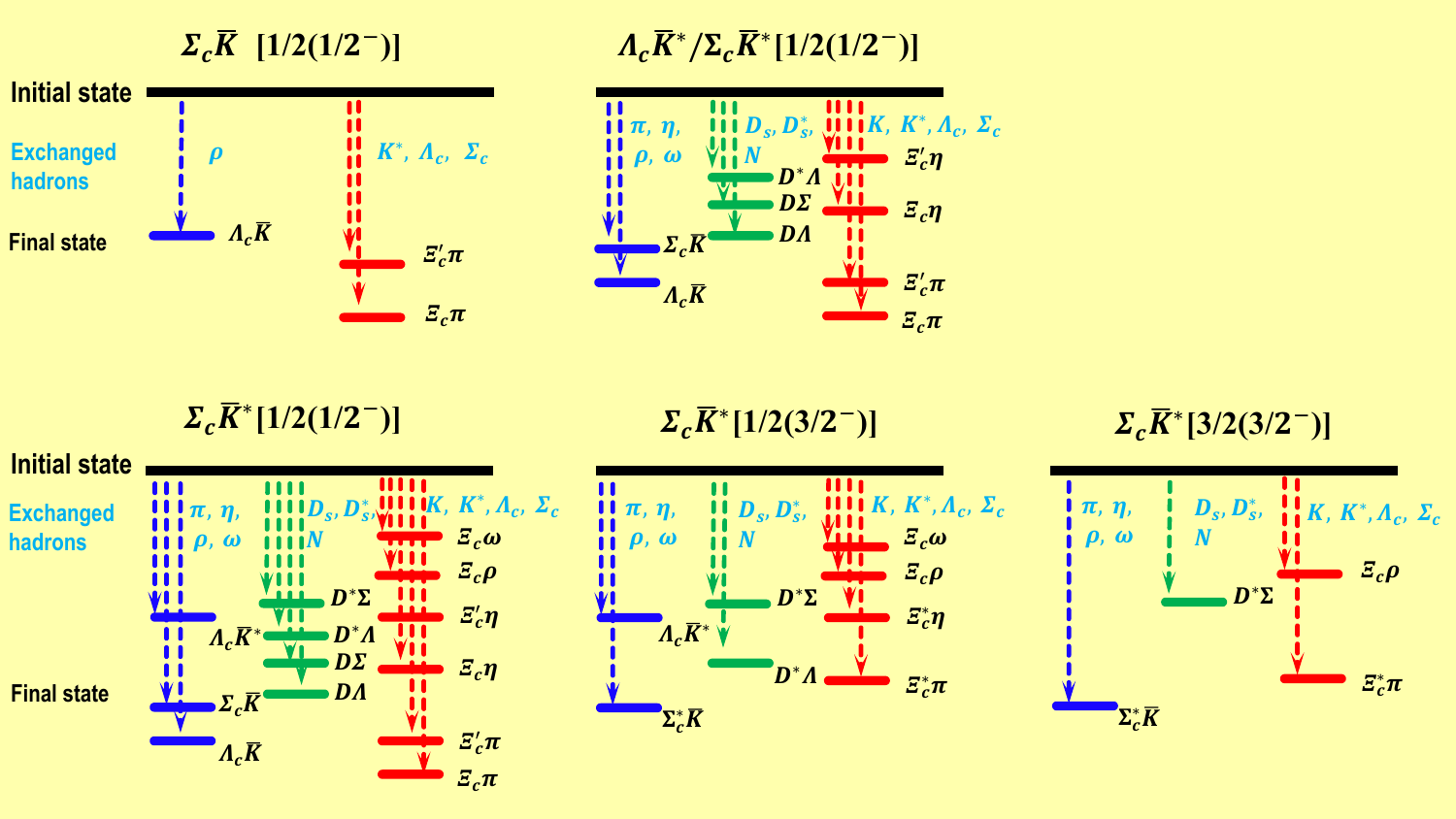}\\
        \includegraphics[width=0.9\linewidth]{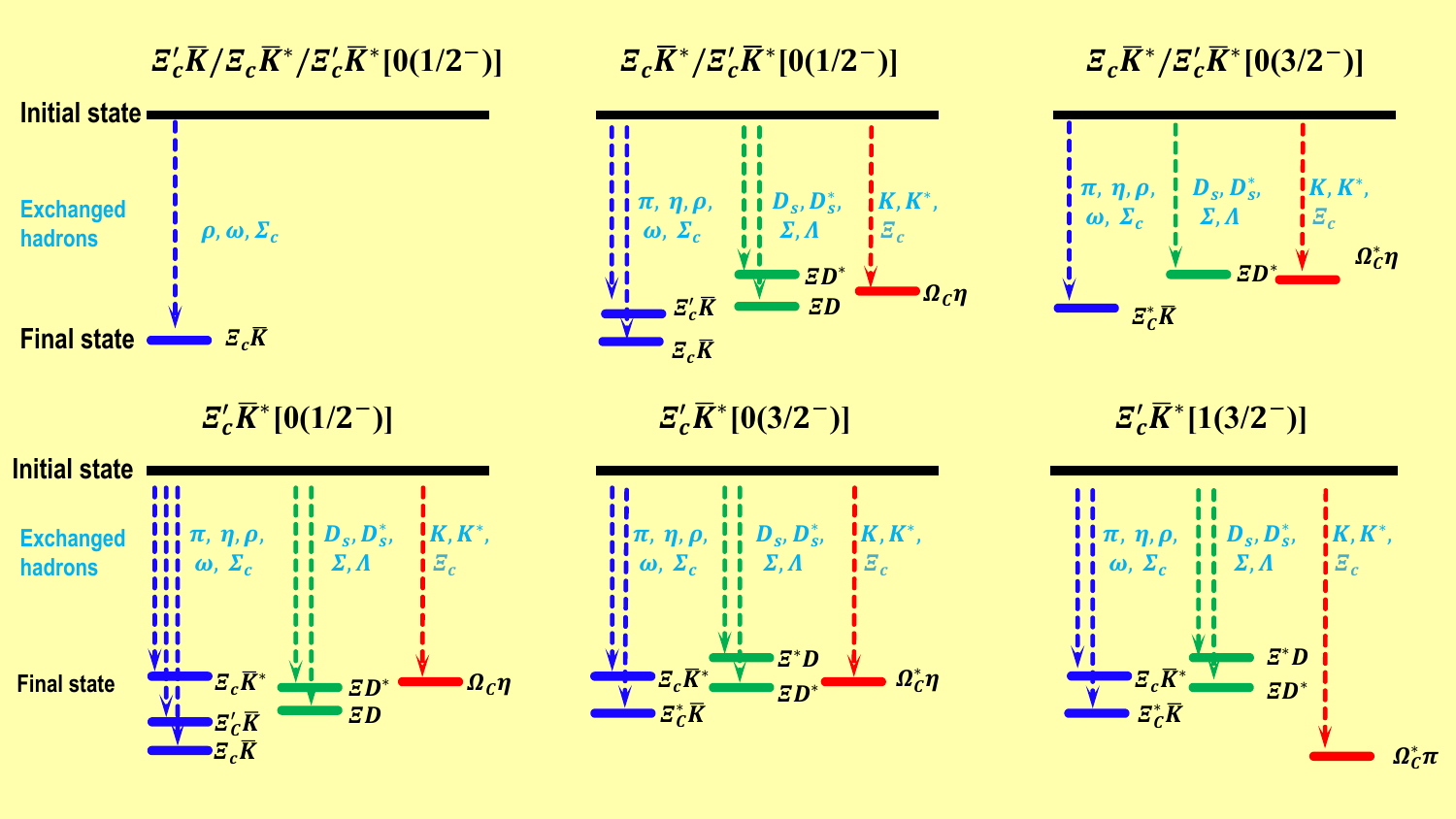}
         \caption{Two-body strong decay channels for $Y_c\bar{K}^{(*)}$  molecules through the $S-$wave interactions}
      \label{channels1}
\end{figure*}

Based on parity conservation and available phase space, we identify the relevant two-body strong decay channels, which are presented in Figure \ref{channels1}. We consider processes mediated by the vertices in our effective Lagrangians, which include $BP \to BP$, $BP \to BV$, $BV \to BV$, and $BV \to DP$. The corresponding analytical expressions for these decay amplitudes are compiled in Table \ref{decayamplitude}.

\begin{table*}[!htbp]
\centering
\caption{Decay amplitudes for $Y_c\bar{K}^{(*)}$ molecular states, where $\mathcal{G}_{I}$ is the isospin factor: $3$ for the isoscalar system, and $-1$ for the isovector system.}
\label{decayamplitude}
\renewcommand\tabcolsep{0.3cm}
\renewcommand{\arraystretch}{1.7}
\begin{tabular}{c|ll|ll}
\toprule[1pt]\toprule[1pt]

\textbf{Types}   & \textbf{Processes}     & \textbf{Amplitude relations} & \textbf{Processes}    & \textbf{Amplitude relations}\\
\hline
$BP\to BP$   &$\Sigma_c\bar K\to\Lambda_c\bar{K}$    &$\sqrt{3}\mathcal{M}^\rho$ &$\Sigma_c\bar K\to\Xi_c^{(\prime)}\pi$   
&$\mathcal{M}^{K^{*}} +2\mathcal{M}^{\Sigma_{c}} -\frac{1}{3}\mathcal{M}^{\Lambda_{c}}$ 
\\
    &$\Xi_c^{\prime}\bar{K}\to \Xi_{c}\bar{K}$ &$\frac{\mathcal{G}_{I}}{\sqrt{2}}\mathcal{M}^{\rho} +\frac{1}{\sqrt{2}}\mathcal{M}^{\omega}+\mathcal{M}^{\Sigma_{c}}$   
    
& \\\hline

$BV\to BP$ &$\Sigma_c \bar{K}^{*}\to\Lambda_c\bar{K}$  
&$\sqrt{3}\mathcal{M}^{\pi} +\sqrt{3}\mathcal{M}^{\rho}$  &$\Sigma_c \bar{K}^{*}\to\Sigma_c\bar{K}$ 
&$2\mathcal{M}^{\pi} +\mathcal{M}^{\eta}+2\mathcal{M}^{\rho} +\mathcal{M}^{\omega}$
\\

 &$\Sigma_c \bar{K}^{*}\to D\Lambda$  &$\sqrt{3}\mathcal{M}^{N} $ &$\Sigma_c \bar{K}^{*}\to D\Sigma$ 
&$-\mathcal{M}^{N} +\mathcal{M}^{D_{s}} +\mathcal{M}^{D_{s}^{*}}$\\

  &$\Sigma_c \bar{K}^{*}\to\Xi_c^{(\prime)}\pi$      
 &$\mathcal{M}^{K}+
\mathcal{M}^{K^{*}}+2\mathcal{M}^{\Sigma_{c}}-\frac{1}{3}\mathcal{M}^{\Lambda_{c}}$   &$\Sigma_c \bar{K}^{*}\to\Xi_c^{(\prime)}\eta$      
&$\sqrt{3}\mathcal{M}^{K}+\sqrt{3}\mathcal{M}^{K^{*}}+\sqrt{3}\mathcal{M}^{\Sigma_{c}}$\\

 &$\Lambda_c \bar{K}^{*}\to\Lambda_c\bar{K}$  &$\mathcal{M}^{\eta} +\mathcal{M}^{\omega}$
  &$\Lambda_c \bar{K}^{*}\to\Sigma_c\bar{K}$ 
&$\sqrt{3}\mathcal{M}^{\pi} +\sqrt{3}\mathcal{M}^{\rho}$\\

&$\Lambda_c\bar{K}^{*}\to D\Lambda$  &$\mathcal{M}^{N} +\mathcal{M}^{D_{s}} +\mathcal{M}^{D_{s}^{*}}$    &$\Lambda_c \bar{K}^{*}\to D\Sigma$ &$\sqrt{3}\mathcal{M}^{N}$\\

 &$\Lambda_c \bar{K}^{*}\to\Xi_c^{(\prime)}\pi$      
 &$-\sqrt{3}\mathcal{M}^{K}-\sqrt{3}\mathcal{M}^{K^{*}}-\sqrt{3}\mathcal{M}^{\Sigma_{c}}$

  &$\Lambda_c \bar{K}^{*}\to\Xi_c^{(\prime)}\eta$  &$\mathcal{M}^{K}+\mathcal{M}^{K^{*}}+\mathcal{M}^{\Lambda_c }$\\

& $\Xi_{c}^{(\prime)}\bar{K}^{*} \to \Xi D$ &$\mathcal{M}^{D_{s}}+\mathcal{M}^{D_{s}^{*}} +\mathcal{M}^{\Sigma}+\mathcal{M}^{\Lambda}$
&$ \Xi_c^{(\prime)}\bar{K}^{*} \to \Xi_c^{(\prime)} \bar{K}$
&$\frac{\mathcal{G}_{I}}{\sqrt{2}}\mathcal{M}^{\pi} +\frac{1}{\sqrt{2}}\mathcal{M}^{\eta}+\frac{\mathcal{G}_{I}}{\sqrt{2}}\mathcal{M}^{\rho} +\frac{1}{\sqrt{2}}\mathcal{M}^{\omega}+\mathcal{G}_{I}\mathcal{M}^{\Sigma_{c}}$\\

 & $\Xi_{c}^{(\prime)}\bar{K}^{*} \to \Omega_{c}\eta$ &$\sqrt{2}\mathcal{M}^{K}+\sqrt{2}\mathcal{M}^{K^{*}} +\mathcal{M}^{\Xi_{c}}$ &\\\hline

$BV\to BV$ &$\Sigma_c \bar{K}^{*}\to D^{*}\Lambda$  &$\sqrt{3}\mathcal{M}^{N} $  &$\Sigma_c \bar{K}^{*}\to D^{*}\Sigma$  
&$-\mathcal{M}^{N} +\mathcal{M}^{D_{s}} +\mathcal{M}^{D_{s}^{*}}$\\

 &$\Sigma_c \bar{K}^{*}\to \Xi_{c}\rho$ 
&$\mathcal{M}^{K}+\mathcal{M}^{K^{*}}+2\mathcal{M}^{\Sigma_{c}}-\frac{1}{3}\mathcal{M}^{\Lambda_{c}}$ &$\Sigma_c \bar{K}^{*}\to \Xi_{c}\omega$ 
&$\sqrt{3}\mathcal{M}^{K}+\sqrt{3}\mathcal{M}^{K^{*}}+\sqrt{3}\mathcal{M}^{\Sigma_{c}}$\\

 &$\Sigma_c \bar{K}^{*}\to \Lambda_c \bar{K}^{*}$  &$\sqrt{3}\mathcal{M}^{\pi} +\sqrt{3}\mathcal{M}^{\rho}$
 &$\Lambda_c \bar{K}^{*}\to D^{*}\Sigma$  
&$\mathcal{M}^{N} +\mathcal{M}^{D_{s}} +\mathcal{M}^{D_{s}^{*}}$ \\

 &$\Xi_{c}^{(\prime)}\bar{K}^{*} \to \Xi D^{*}$ &$\mathcal{M}^{D_{s}}+\mathcal{M}^{D_{s}^{*}} +\mathcal{M}^{\Sigma}+\mathcal{M}^{\Lambda}$
 &$\Xi_{c}^{\prime}\bar{K}^{*} \to \Xi_{c}^{}\bar{K}^{*}$ &$\frac{\mathcal{G}_{I}}{\sqrt{2}}\mathcal{M}^{\pi} +\frac{1}{\sqrt{2}}\mathcal{M}^{\eta}+\frac{\mathcal{G}_{I}}{\sqrt{2}}\mathcal{M}^{\rho} +\frac{1}{\sqrt{2}}\mathcal{M}^{\omega}+\mathcal{G}_{I}\mathcal{M}^{\Sigma_{c}}$
 \\\hline

$BV\to DP$ 
 &$\Sigma_c \bar{K}^{*}\to \Xi_{c}^{*}\pi$ 
 &$\mathcal{M}^{K}+\mathcal{M}^{K^{*}}+2\mathcal{M}^{\Sigma_{c}}-\frac{1}{3}\mathcal{M}^{\Lambda_{c}}$
  &$\Sigma_c \bar{K}^{*}\to \Xi_{c}^{*}\eta$ 
&$\sqrt{3}\mathcal{M}^{K}+\sqrt{3}\mathcal{M}^{K^{*}}+\sqrt{3}\mathcal{M}^{\Sigma_{c}}$\\

 &$\Sigma_c \bar{K}^{*}\to \Sigma_{c}^{*}\bar{K}$ 
&$\mathcal{M}^{\pi} -\mathcal{M}^{\eta}+\mathcal{M}^{\rho} -\mathcal{M}^{\omega}$ &$\Xi_{c}^{\prime}\bar{K}^{*} \to \Xi^{*}D $  
&$\mathcal{M}^{D_{s}}+\mathcal{M}^{D_{s}^{*}} +\sqrt{2}\mathcal{M}^{\Sigma}+\mathcal{M}^{\Lambda}$
 \\

& $\Xi_{c}^{(\prime)}\bar{K}^{*} \to \Omega_{c}^{*}\eta$ &$\sqrt{2}\mathcal{M}^{K}+\sqrt{2}\mathcal{M}^{K^{*}}$
 &$\Xi_{c}^{\prime}\bar{K}^{*} \to \Xi_{c}^{*}\bar{K}$  
&$\frac{\mathcal{G}_{I}}{\sqrt{2}}\mathcal{M}^{\pi} +\frac{1}{\sqrt{2}}\mathcal{M}^{\eta}+\frac{\mathcal{G}_{I}}{\sqrt{2}}\mathcal{M}^{\rho} +\frac{1}{\sqrt{2}}\mathcal{M}^{\omega}+\mathcal{G}_{I}\mathcal{M}^{\Sigma_{c}}$\\

& $\Xi_{c}^{(\prime)}\bar{K}^{*} \to \Omega_{c}^{*}\pi$ &$\sqrt{2}\mathcal{M}^{K}+\sqrt{2}\mathcal{M}^{K^{*}}+\mathcal{M}^{\Xi_c}$
 &\\

\midrule[1pt]\midrule[1pt]

\multicolumn{5}{l}{$\mathcal{M}_{BP\to BP}^{V}=\left\{g_{BBV}\bar{u}_3\gamma^{\mu}u_1
      +\frac{f_{BBV}}{4m_B}\bar{u}_3(\gamma^{\mu}\gamma^{\nu}-\gamma^{\nu}\gamma^{\mu})q_{\nu}u_1\right\}\frac{g_{\mu\beta}-q_{\mu}q_{\beta}/m_{V}^2}{q^2-m_{V}^2}\left\{-g_{PPV}(p_4^{\beta}+p_2^{\beta})\right\}$}\\
      
\multicolumn{5}{l}{$\mathcal{M}_{{BP\to BV}}^{{V}} = \left\{g_{BBV}\bar{u}_3\gamma^{\mu}u_1+\frac{f_{BBV}}{4m_{B}}\bar{u}_3(\gamma^{\mu}\gamma^{\nu}
-\gamma^{\nu}\gamma^{\mu})q_{\nu}u_1\right\}\frac{g_{\mu\beta}-q_{\mu}q_{\beta}/m_{V}^2}{q^2-m_{V}^2}\frac{g_{VVP}}{m_V}\varepsilon^{\lambda\nu\alpha\beta}p_{4\nu}\epsilon_{4\lambda}^{\dag}q_{\alpha}$}\\
      
\multicolumn{5}{l}{$\mathcal{M}_{{BP\to VB}}^{{B}} = \frac{g_{BBP}}{m_P}\bar{u}_4\gamma^\mu p_{2\mu}\gamma_5\rlap\slash{q}\frac{1}{\rlap\slash{q}-m_{B}}
     \left\{g_{BBV}\epsilon_{3\mu}^{\dag}\gamma^{\mu}u_1  -\frac{f_{BBV}}{4m_{B}}p_{3\mu}\epsilon_{3\nu}^{\dag}
(\gamma^{\mu}\gamma^{\nu}-\gamma^{\nu}\gamma^{\mu})q_{\nu}u_1\right\}$}\\
      
\multicolumn{5}{l}{$\mathcal{M}_{{BV\to BV}}^{{P}} = -\frac{g_{BBP}}{m_P }\bar{u}_3\gamma_5\gamma^\mu q_\mu u_1\frac{1}{q^2-m_{P}^2}\frac{g_{VVP}}{m_V}\varepsilon^{\lambda\sigma\alpha\beta} p_{4\lambda}\epsilon_{4\sigma}^{\dag}p_{2\alpha}\epsilon_{2\beta}$}\\
      
\multicolumn{5}{l}{$\mathcal{M}_{{BV\to BV}}^{{V}} = \left\{g_{BBV}\bar{u}_3\gamma^{\mu}u_1
      +\frac{f_{BBV}}{4m_B}\bar{u}_3(\gamma^{\mu}\gamma^{\nu}-\gamma^{\nu}\gamma^{\mu})q_{\nu}u_1\right\}\frac{g_{\mu\beta}-q_{\mu}q_{\beta}/m_{V}^2}{q^2-m_{V}^2} g_{VVV}\left\{\epsilon_{4}^{\alpha\dag}\epsilon_{2}^{\beta}(p_{2\alpha}-q_{\alpha})\right.\left.
      -\epsilon_{2\alpha}\epsilon_4^{\alpha\dag}(p_2^{\beta}+p_4^{\beta})
      +\epsilon_{2\alpha}(\epsilon_4^{\beta\dag}q^{\alpha}+p_4^{\alpha}\epsilon_4^{\beta\dag})\right\}$}\\
      
\multicolumn{5}{l}{$\mathcal{M}_{{BV\to VB}}^{{B}} = \left\{g_{BBV}\bar{u}_4\gamma^{\mu}\epsilon_{2\mu}
      +\frac{f_{BBV}}{4m_B}\bar{u}_4(\gamma^{\mu}\gamma^{\nu}-\gamma^{\nu}\gamma^{\mu})p_{2\mu}\epsilon_{2\nu}\right\}\frac{1}{\rlap\slash{q}-m_{B}}\left\{g^\prime_{BBV}\gamma^{\alpha}\epsilon_{3\alpha}^{\dag}u_1\right.
      \left.+\frac{f_{BBV}'}{4m'}(\gamma^{\alpha}\gamma^{\beta}-\gamma^{\alpha}\gamma^{\beta})p_{3\alpha}\epsilon_{3\beta}^{\dag}u_1\right\}$}\\
      
\multicolumn{3}{l}{$\mathcal{M}_{{BV\to DP}}^{{P}} = \frac{g_{BDP}}{m_P}\bar{u}_3q_{\mu}u_1^{\mu}\frac{1}{q^2-m_{\mathbb{P}}^2}ig_{PPV}\epsilon_4^{\nu\dag}(q_{\nu}-p_{4\nu})$}

&\multicolumn{2}{l}{$\mathcal{M}_{{BV\to PD}}^{{B}} = i\frac{g_{BBP}}{m_P}\bar{u}_4\gamma^5 \gamma_\alpha p_{2}^\alpha \frac{1}{\rlap\slash{q}-m_{B}} \frac{g_{BDV}}{m_V} \gamma^5(\gamma^{\nu}u_{1}^{\mu}-\gamma^{\mu}u_1^{\nu})q_{\mu}\epsilon_{3\nu}^{\dag}$}\\

\multicolumn{3}{l}{$\mathcal{M}_{{BV\to DP}}^{{V}} = -i\frac{g_{BDV}}{m_V}\bar{u}_3\gamma^5(\gamma^{\nu}u_{1}^{\mu}-\gamma^{\mu}u_1^{\nu})q_{\mu}\frac{g_{\nu\beta}-q_{\nu}q_{\beta}/m_{V}^2}{q^2-m_{V}^2}\frac{g_{VVP}}{m_V}\varepsilon^{\lambda\beta\alpha\delta}q_{\lambda}p_{4\alpha}\epsilon_{4\delta}^{\dag}$}

&\multicolumn{2}{l}{$\mathcal{M}_{{BV\to PD}}^{{B}} = i\frac{g_{BBP}}{m_P}\bar{u}_4\gamma^5 \gamma_\alpha p_{2}^\alpha \frac{1}{\rlap\slash{q}-m_{B}} \frac{g_{BDV}}{m_V} \gamma^5(\gamma^{\nu}u_{1}^{\mu}-\gamma^{\mu}u_1^{\nu})q_{\mu}\epsilon_{3\nu}^{\dag}$}\\

\multicolumn{3}{l}{$\mathcal {M}_{BP \to BP}^{B} = \frac{g_{BBP}}{m_P}\bar{u}_3\gamma^\alpha p_{1\alpha}\gamma_5\frac{1}{\rlap\slash{q}-m_{B}}\frac{g^\prime_{BBP}}{m^\prime_P}\gamma^\mu p_{4\mu}\gamma_5u_2$}

&\multicolumn{2}{l}{$\mathcal{M}_{{BP\to BV}}^{{P}} = \frac{g_{BBP}}{m_P}\bar{u}_3\gamma_5 \rlap\slash{q} u_1\frac{{1}}{q^2-m_{B}^2}g_{PPV}\epsilon_4^{\mu\dag}(q_{\mu}-p_{2\mu})
$}\\

\bottomrule[1pt]\bottomrule[1pt]
\end{tabular}
\end{table*}

\section{Numerical results}\label{sec3}

\subsection{$\Lambda_c\bar{K}^{(*)}$ and $\Sigma_c\bar{K}^{(*)}$ molecular pentaquarks}

\begin{figure*}[!htbp]
     \raggedright
        \includegraphics[width=0.32\linewidth]{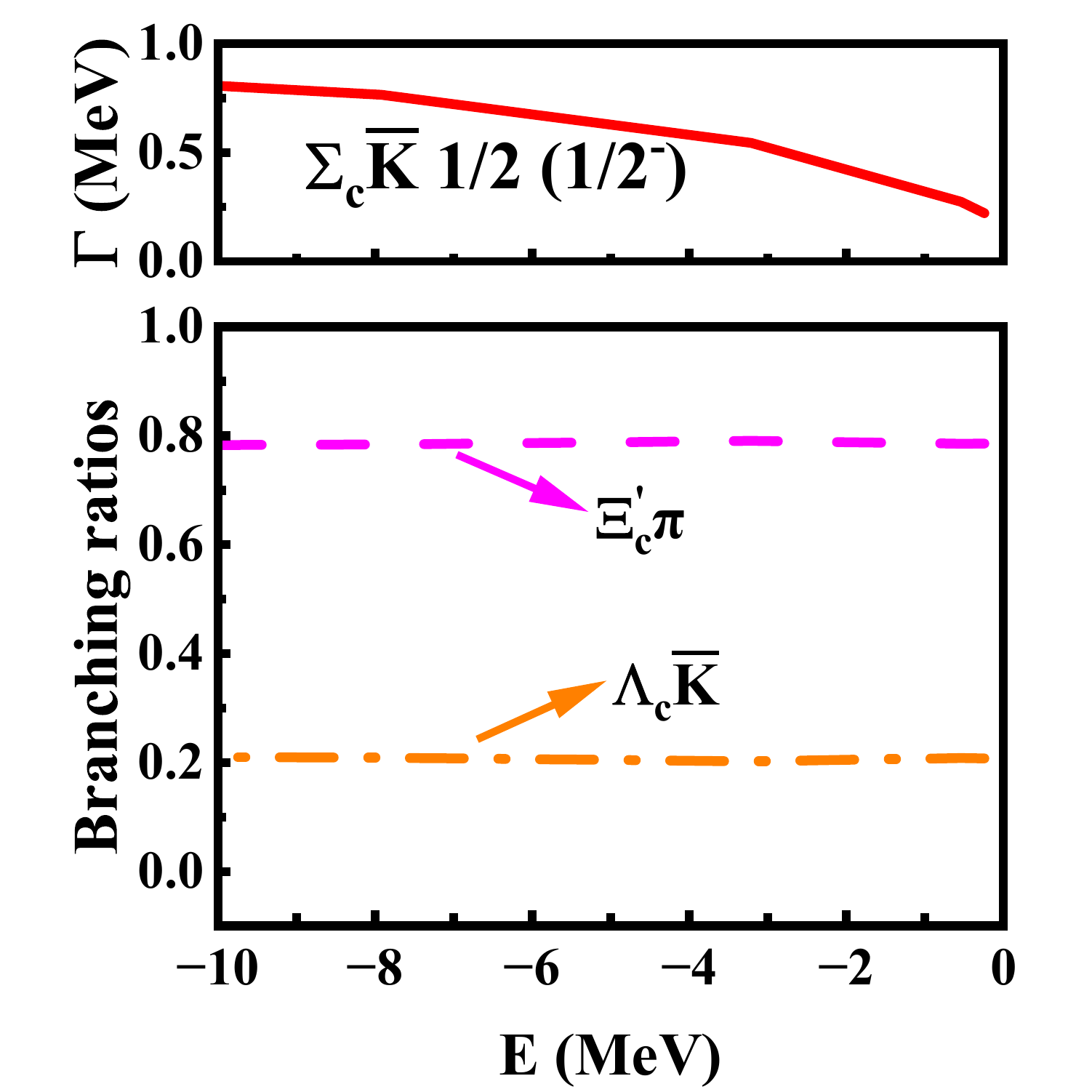}
        \includegraphics[width=0.32\linewidth]{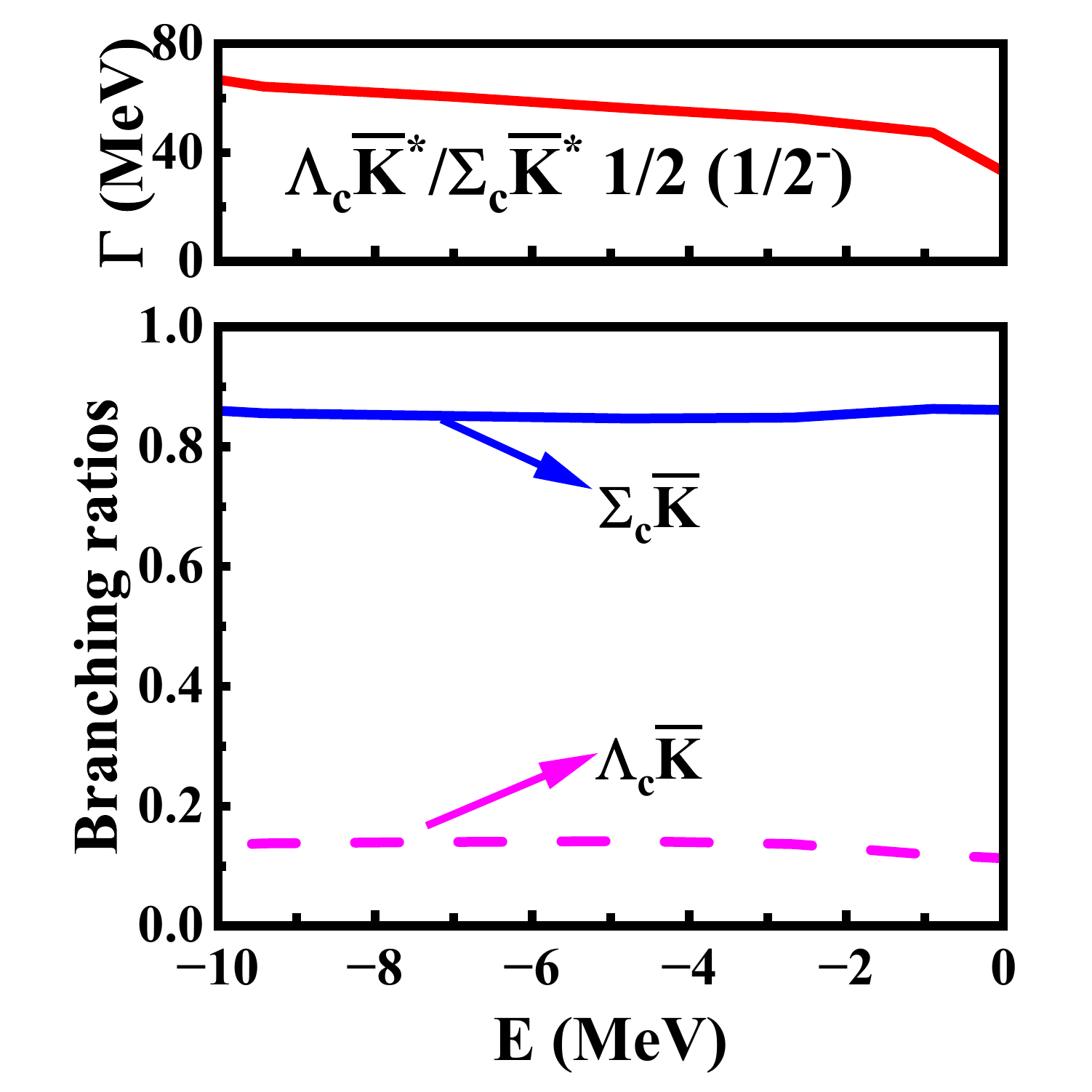}\\
        \includegraphics[width=0.32\linewidth]{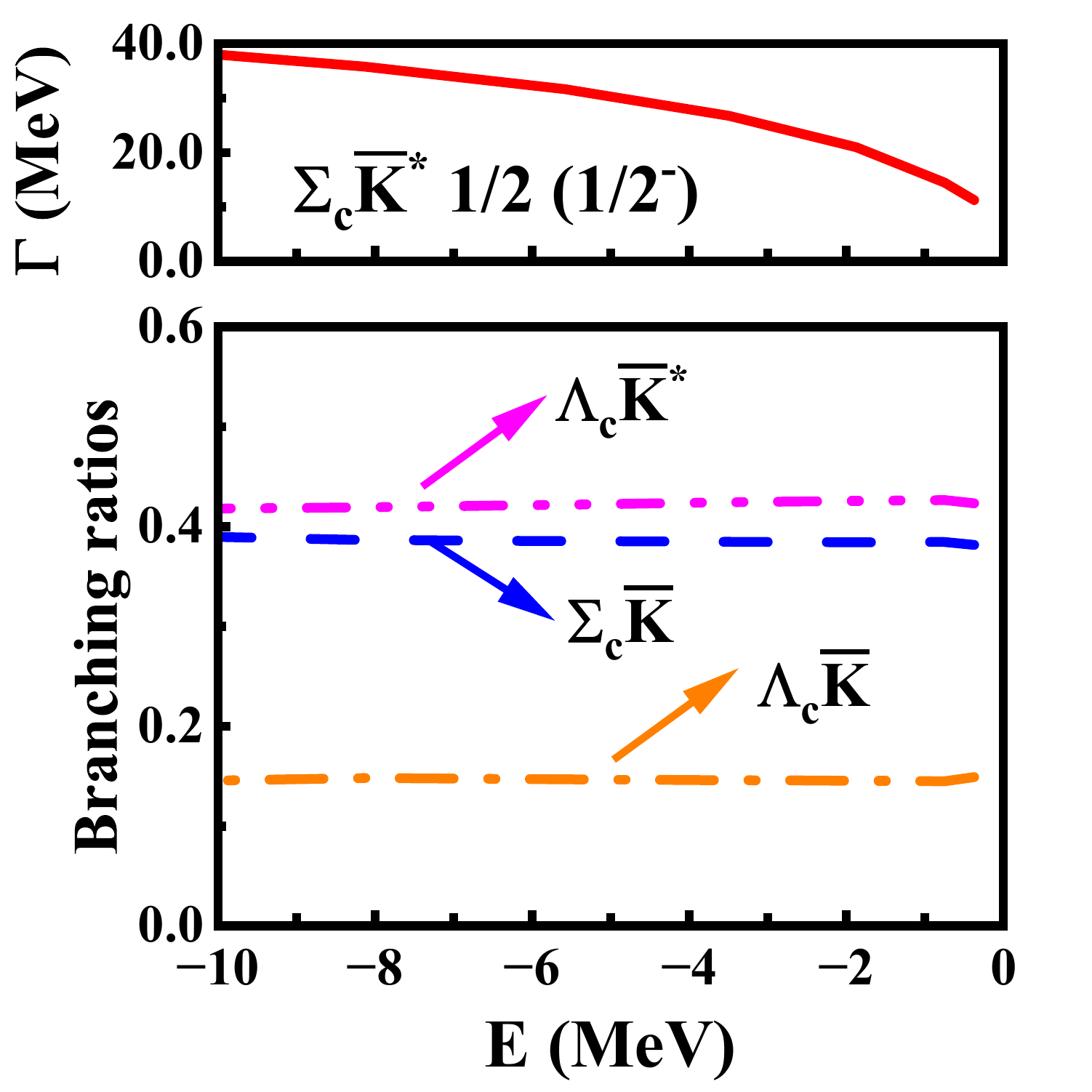}
        \includegraphics[width=0.32\linewidth]{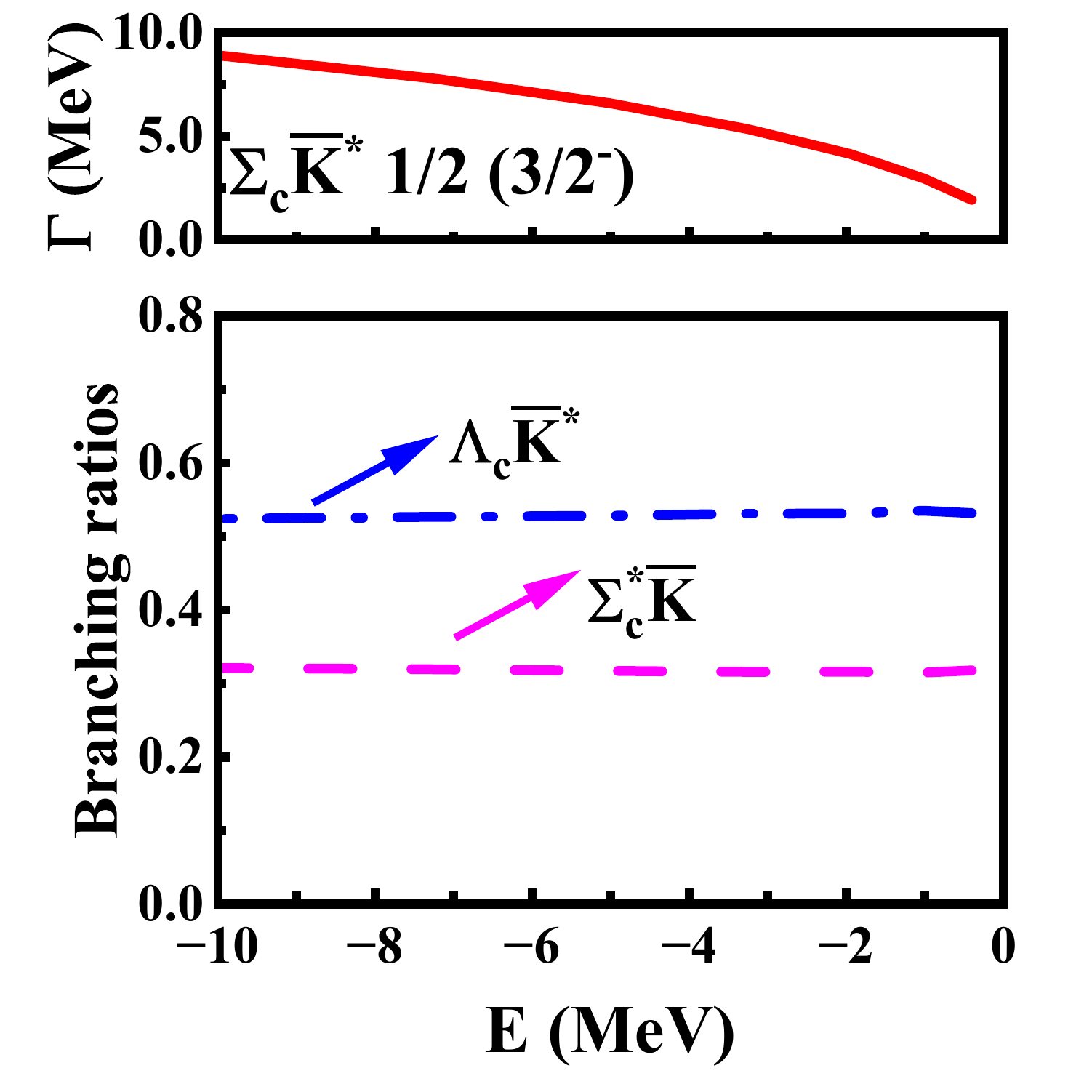}
        \includegraphics[width=0.32\linewidth]{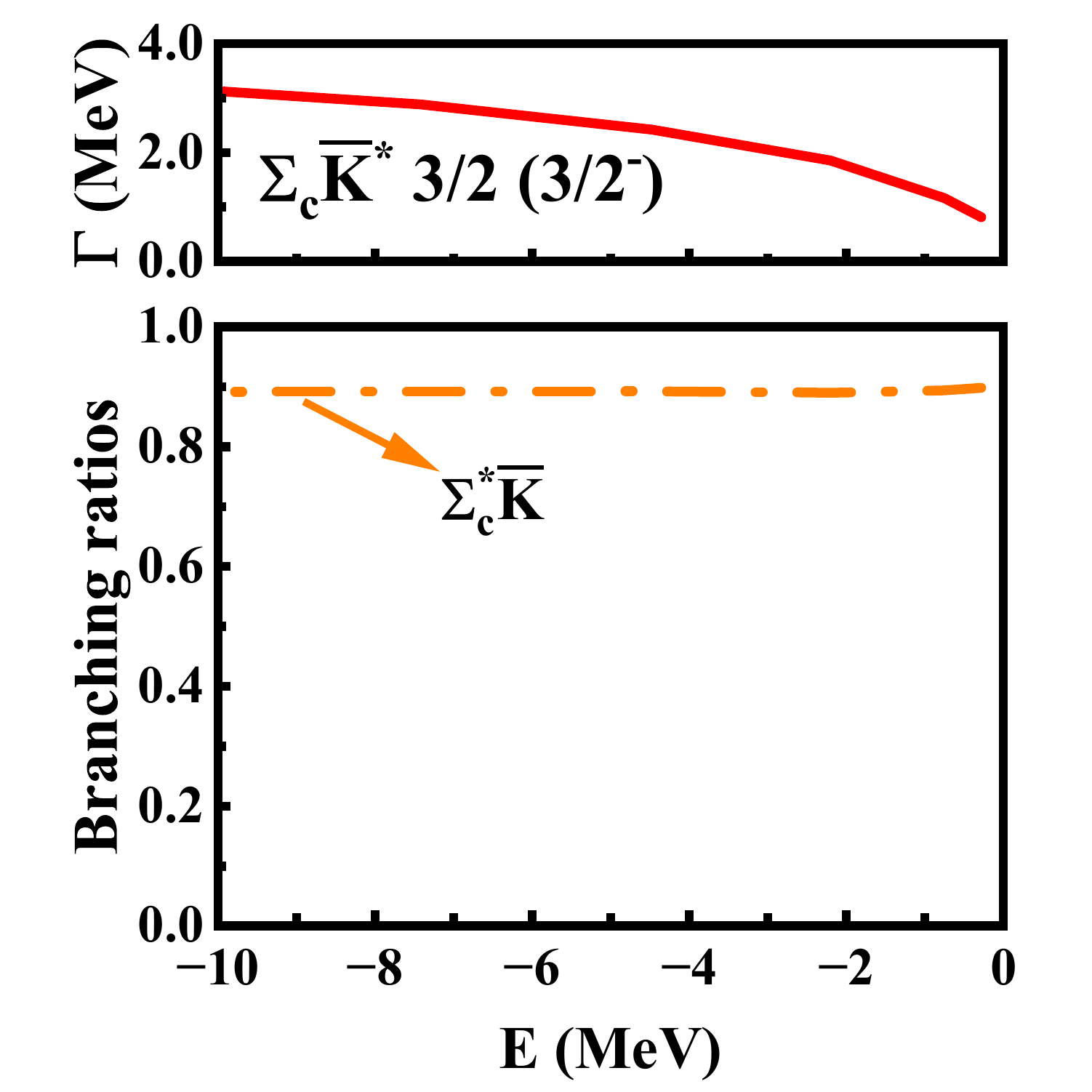}
         \caption{Two-body strong decay behavior for $\Lambda_c \bar{K}^{*}$ and $\Sigma_c \bar{K}^{*}$  molecules through the $S-$wave interactions, and final states with negligible branching ratios are not shown.}
      \label{num1}
\end{figure*}

In this section, we use the previously obtained wave functions \cite{Chen:2023qlx} to compute the partial and total decay widths for possible $\Lambda_c\bar{K}^{(*)}$ and $\Sigma_c\bar{K}^{(*)}$ molecular pentaquarks. In Figure \ref{num1}, we present the corresponding total decay widths and branching ratios. For the $\Sigma_c\bar{K}$ molecule with $1/2(1/2^-)$, the allowed $S$-wave decay channels are $\Lambda_c\bar{K}$, $\Xi_c\pi$, and $\Xi_c^{\prime}\pi$. For binding energies $E > -10$ MeV, the total decay width is less than 1.00 MeV. The decay to $\Xi_c^{\prime}\pi$ is dominant, with a branching ratio of approximately 78\%, and proceeds primarily via $K$-meson exchange. The partial widths to $\Lambda_c\bar{K}$ and $\Xi_c\pi$ are on the order of several tens of keV and several hundreds of keV, respectively. 

As established in Ref. \cite{Wang:2022oof}, coupled-channel effects are crucial for the formation of the $\Lambda_c\bar{K}^*/\Sigma_c\bar{K}^*$ coupled molecule with $1/2(1/2^-)$, at a binding energy of approximately $-10$ MeV, the $\Sigma_c\bar{K}^*$ component constitutes about 40\% of the wave function. The considered decay channels include $\Lambda_c\bar{K}$, $\Sigma_c\bar{K}$, $D^{(*)}\Lambda$, $D\Sigma$, $\Xi_c^{( \prime)}\pi$, and $\Xi_c^{(\prime)}\eta$. Using the coupled-channel wave functions from Ref. \cite{Chen:2023qlx}, we find total decay widths in the range of 30–76 MeV for $E > -10$ MeV. The $\Sigma_c\bar{K}$ channel dominates, with a branching ratio of about 85\%, due to the strong pion-exchange mechanism. The $\Lambda_c\bar{K}$ channel is secondary, with a branching ratio of $\sim 14\%$, also facilitated by pion exchange after considering the coupled channel effects.

For the $\Sigma_c\bar{K}^*$ molecule with $1/2(1/2^-)$, the total decay width is on the order of several tens of MeV for binding energy $E>-10$ MeV. The dominant channels are $\Lambda_c\bar{K}^*$, $\Sigma_c\bar{K}$, and $\Lambda_c\bar{K}$, with branching ratios of approximately 42\%, 38\%, and 15\%, respectively. Their prominence stems from the presence of strong light meson (especially pion) exchange diagrams. For the remaining decay modes, we find this state exhibits a clear preference for decaying into a charmed-strange baryon and a light meson (e.g., $\Xi_c\pi$) rather than a charmed-strange meson and a hyperon (e.g., $D\Sigma$). The former process benefits from strange meson exchange, while the latter involves the exchange of a nucleon or a charmed-strange meson, leading to suppressed amplitudes.

For the $\Sigma_c\bar{K}^*$ molecule with $1/2(3/2^-)$, the total decay width for this state is several times smaller than that of its $1/2(1/2^-)$ counterpart, amounting to several MeV for $E > -10$ MeV. The dominant decay channel is $\Lambda_c\bar{K}^*$ (branching ratio $\sim 53\%$), followed by $\Sigma_c^*\bar{K}$ ($\sim 32\%$). Both processes are enhanced by light meson (pion) exchange.

For the $\Sigma_c\bar{K}^*$ molecule with $3/2(3/2^-)$, its two-body strong decay width is several times smaller than the $1/2(3/2^-)$ state, remaining at the level of several MeV for $E > -10$ MeV. The decay is overwhelmingly dominated by the $\Sigma_c^*\bar{K}$ channel, with a branching ratio of approximately 90\%, again driven by efficient pion exchange. The remaining decay channels have negligible branching fractions.

\subsection{Charmed molecular pentaquarks with strangeness $|S|=2$}

\begin{figure*}[!htbp]
     \raggedright
        \includegraphics[width=0.32\linewidth]{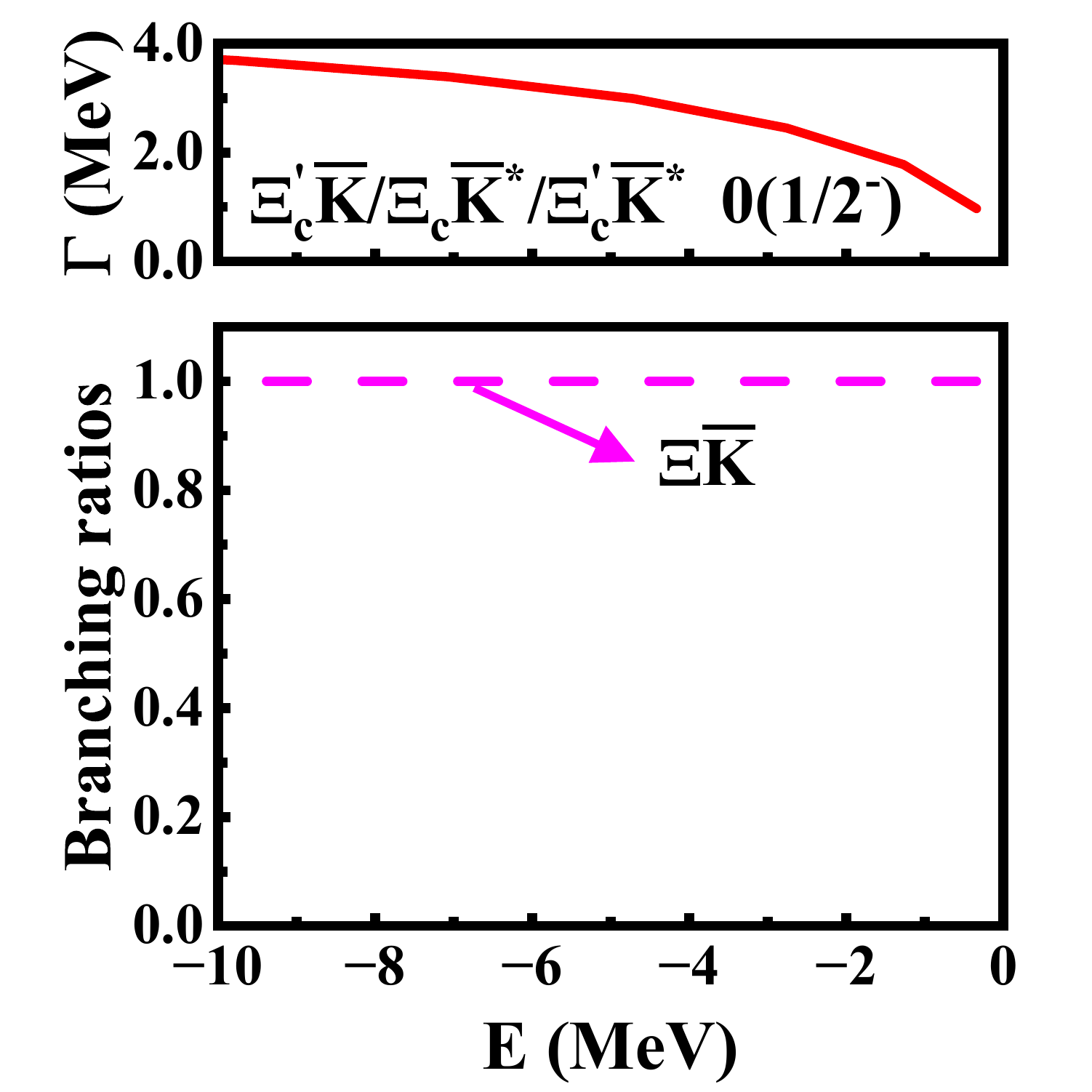}
        \includegraphics[width=0.32\linewidth]{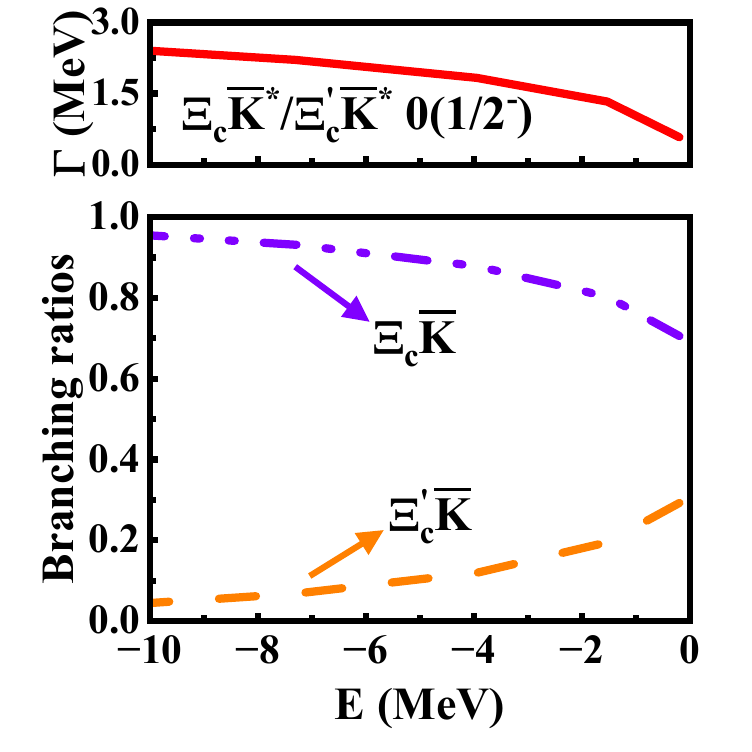}
        \includegraphics[width=0.32\linewidth]{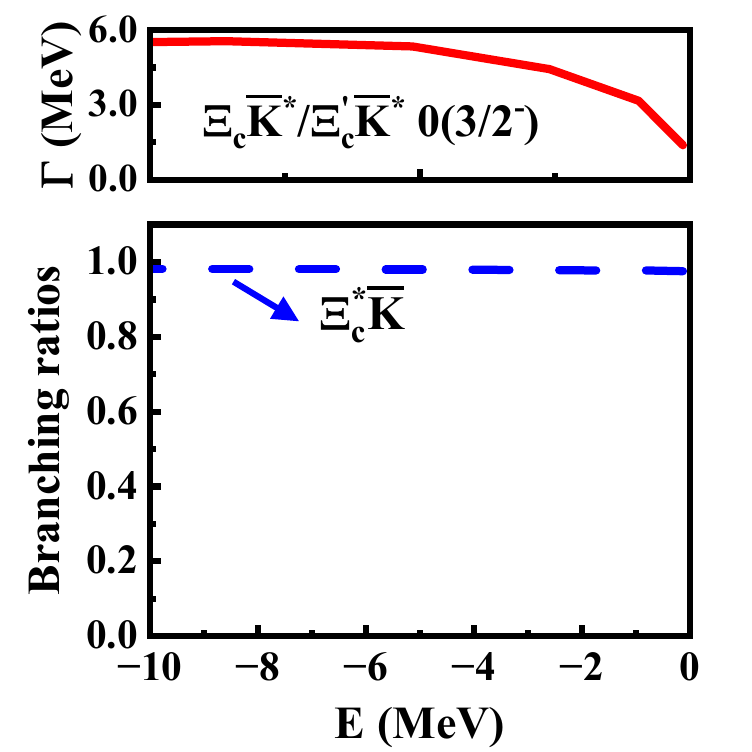}\\
        \includegraphics[width=0.32\linewidth]{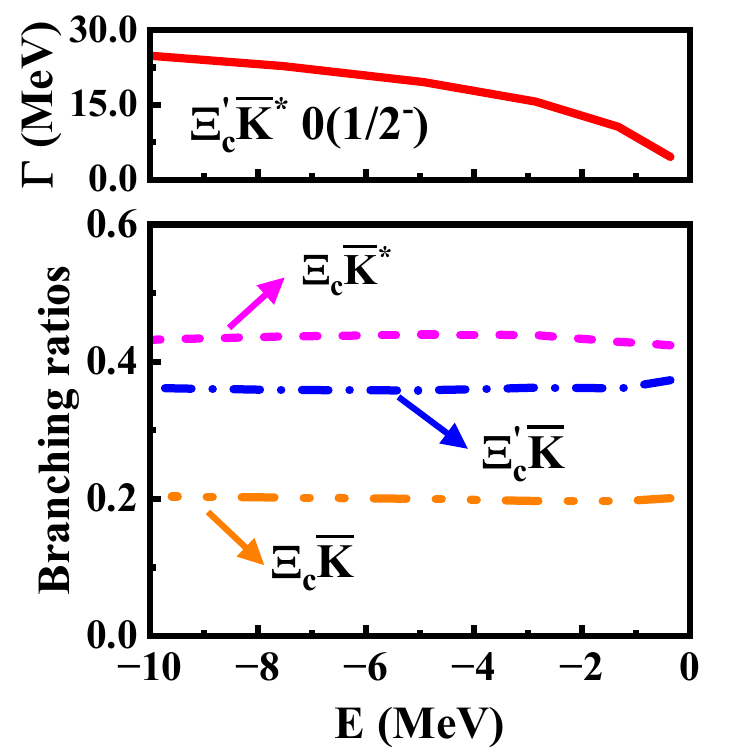}
        \includegraphics[width=0.32\linewidth]{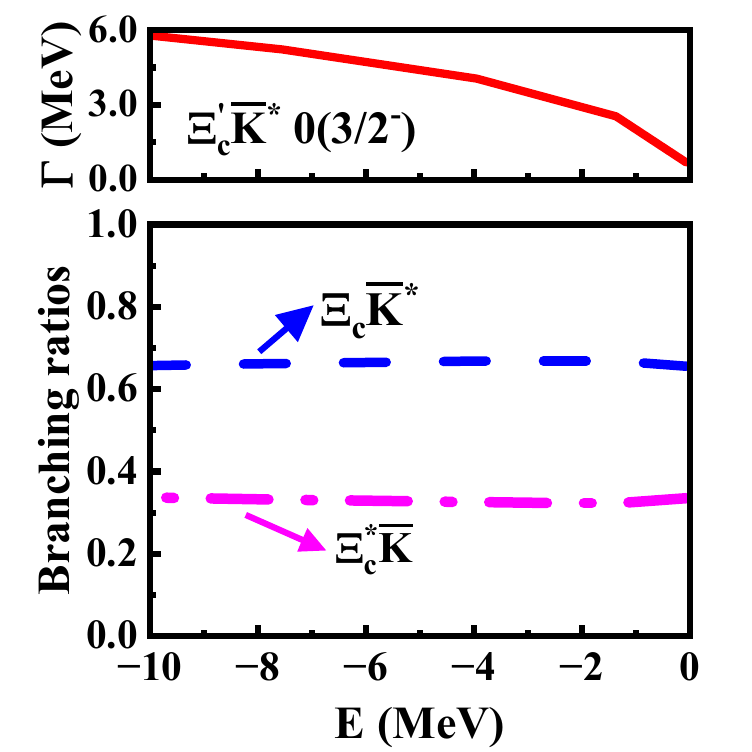}
        \includegraphics[width=0.32\linewidth]{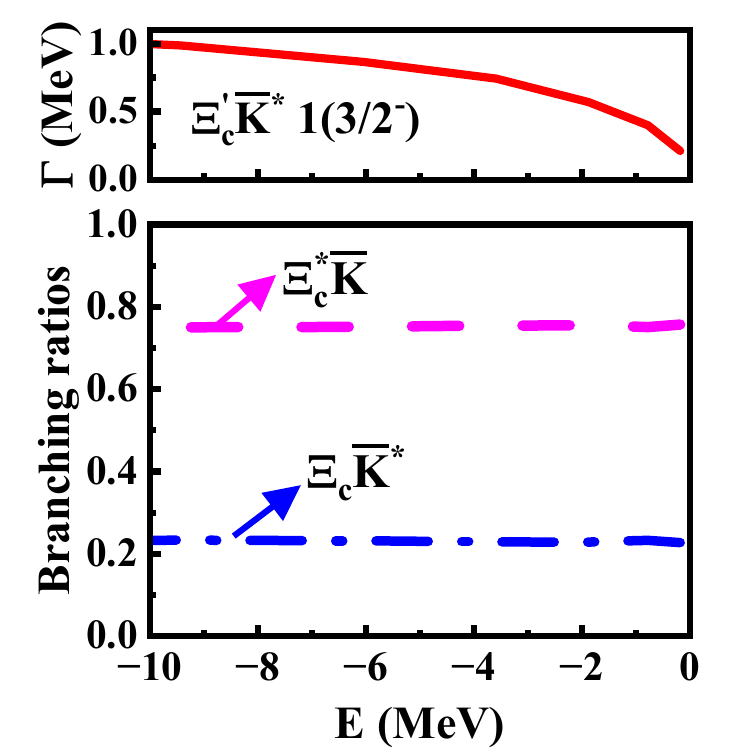}
         \caption{Two-body strong decay behavior for $\Xi_c^{(\prime)} \bar{K}^{*}$  molecules through the $S-$wave interactions, and final states with negligible branching ratios are not shown.}
      \label{num2}
\end{figure*}

In Figure \ref{num2}, we present the two-body strong decay properties for the predicted $\Xi_c^{(\prime)}\bar{K}^{(*)}$ molecular pentaquarks. The key findings are summarized as follows:
\begin{itemize}
\item $\Xi_c^{\prime}\bar{K}/\Xi_c\bar{K}^*/\Xi_c^{\prime}\bar{K}^* $ coupled molecule with $I(J^P)=0(1/2^-)$: The only allowed $S$-wave decay channel is $\Xi_c\bar{K}$. For binding energies $E > -10$ MeV, the corresponding partial width is on the order of several MeV.

\item $\Xi_c\bar{K}^*/\Xi_c^{\prime}\bar{K}^*$ coupled molecule with $I(J^P)=0(1/2^-)$: The accessible $S$-wave decay channels are $\Xi_c^{(\prime)}\bar{K}$, $\Xi D^{(*)}$, and $\Omega_c\eta$. As the binding energy varies from 0 to $-10$ MeV, the total decay width increases from approximately 0.6 MeV to 2.5 MeV. The decay is overwhelmingly dominated by the $\Xi_c\bar{K}$ channel, which accounts for about 90\% of the branching ratio, followed by $\Xi_c^{\prime}\bar{K}$. The dominance of the $\Xi_c\bar{K}$ mode is attributed to the efficient pion-exchange mechanism. The remaining three channels have negligible branching fractions due to the larger masses of the exchanged mesons (e.g., strange mesons or charmed mesons).

\item $\Xi_c\bar{K}^*/\Xi_c^{\prime}\bar{K}^*$ coupled molecule with $I(J^P)=0(3/2^-)$: The allowed $S$-wave decays are $\Xi_c^*\bar{K}$, $\Xi D^*$, and $\Omega_c^*\eta$. For $E > -10$ MeV, the total width ranges from 1.47 MeV to 5.38 MeV. The $\Xi_c^*\bar{K}$ channel is dominant, with a branching ratio close to 98\%, while the contributions from $\Omega_c^*\eta$ and $\Xi D^*$ are negligible. This indicates the pivotal role of the $\Xi_c^*\bar{K}$ component in the decay dynamics. The branching ratios show little dependence on the binding energy.

\item $\Xi_{c}^{\prime} \bar{K}^*$ molecule with $I(J^P)=0(1/2^-)$: The considered decay channels include $\Xi_c^{(\prime)}\bar{K}$, $\Xi_c\bar{K}^*$, $\Xi D^{(*)}$, and $\Omega_c\eta$. For $E > -10$ MeV, the total decay width reaches several tens of MeV. The primary decay modes are $\Xi_c\bar{K}^*$, $\Xi_c^{\prime}\bar{K}$, and $\Xi_c\bar{K}$, with branching ratios of approximately 44\%, 35\%, and 20\%, respectively. Their prominence is due to the strong pion-exchange diagrams. The remaining channels have partial widths of only a few to several tens of keV, resulting in very small branching ratios. Notably, the $\Omega_c\eta$ channel, though allowed via strange meson ($K$, $K^*$) exchange, is highly suppressed due to the inherently small $K^{(*)}K^{(*)}\eta$ couplings.

\item $\Xi_{c}^{\prime} \bar{K}^*$ molecule with $I(J^P)=0(3/2^-)$: The $S$-wave decay channels are $\Xi_c^*\bar{K}$, $\Xi_c\bar{K}^*$, $\Xi D^*$, $\Xi^*D$, and $\Omega_c^*\eta$. The total decay width is on the order of several MeV for $E > -10$ MeV. The dominant channels are $\Xi_c\bar{K}^*$ and $\Xi_c^*\bar{K}$, with branching ratios of 66\% and 33\%, respectively, a consequence of the strong pion-exchange contributions. Among the remaining modes, $\Xi D^*$ has the largest branching ratio.

\item $\Xi_{c}^{\prime} \bar{K}^*$ molecule with $I(J^P)=1(3/2^-)$: Although the decay channels are similar to those of its isoscalar counterpart, the total decay width is suppressed by a factor of a few. This suppression arises from two factors: first, the interaction strength mediated by isovector mesons ($\pi$, $\rho$) is a factor of three smaller for the $I=1$ state compared to the $I=0$ state; second, there is significant destructive interference between the $\rho$-meson and $\omega$-meson exchange amplitudes. For $E > -10$ MeV, the total width is approximately 1 MeV. Due to the surviving pion-exchange contributions, the dominant channels are $\Xi_c^*\bar{K}$ and $\Xi_c\bar{K}^*$, with branching ratios of 75\% and 23\%, respectively.
\end{itemize}

\section{Summary}\label{sec4}

\begin{table*}[!htbp]
\renewcommand\tabcolsep{0.70cm}
\renewcommand{\arraystretch}{1.8}
\caption{A summary of the two-body decay properties for the possible $Y_cK^{(*)}$ molecules. Here, $E$ and $\Gamma$ stands for the range of the binding energy and the total width, respectively. The units of the binding energy $E$ and the decay width $\Gamma$ are MeV.} \label{summary}
\begin{tabular}{lc|c|ccc}
\toprule[1pt]\toprule[1pt]
States &$I(J^P)$  &$\Gamma$   &\multicolumn{3}{c}{ Primary Decay} \\\hline
$\Sigma_c \bar{K}$   &$1/2(1/2^-)$  &$(0.22,0.90)$     
   &$\Xi_c^{\prime} \pi (78\%)$ &$\Lambda_{c}\bar{K}(20\%)$    \\
   
$\Lambda_c \bar{K}^*/\Sigma_c \bar{K}^*$    &$1/2(1/2^-)$    &$(30,76)$ 
&$\Sigma_c \bar{K}(85\%)$   &$\Lambda_c \bar{K}(14\%)$      \\

$\Sigma_c \bar{K}^*$   &$1/2(1/2^-)$   &$(11,40)$  &$\Lambda_c \bar{K}^*(42\%)$  &$\Sigma_c \bar{K}(38\%)$   &$\Lambda_c \bar{K}(15\%)$         \\

                      &$1/2(3/2^-)$ &$(2,9.80)$   &$\Lambda_c \bar{K}^*(53\%)$ &$\Sigma_c^*\bar{K}(32\%)$      & \\
                     &$3/2(3/2^-)$   &$(0.8,3.2)$   &$\Sigma_c^*\bar{K}(90\%)$     \\

$\Xi_c^{\prime} \bar{K}/\Xi_c \bar{K}^*/\Xi_c^{\prime} \bar{K}^*$ &$0(1/2^-)$   &$(0.97,3.9)$     
  &$\Xi_c \bar{K}(100\%)$        \\
$\Xi_c \bar{K}^*/\Xi_c^{\prime} \bar{K}^*$    &$0(1/2^-)$  &$(0.6,2.5)$      
   &$\Xi_c \bar{K}(90\%)$     &$\Xi_c^{\prime} \bar{K}(10\%)$   &    \\
   
               &$0(3/2^-)$   &$(1.47,5.38)$       &$\Xi_c^* \bar{K}(98\%)$ \\  
               
$\Xi_c^{\prime} \bar{K}^*$  &$0(1/2^-)$   &$(4.52,25)$   &$\Xi_c \bar{K}^*(44\%)$  &$\Xi_c^{\prime} \bar{K}(35\%)$   &$\Xi_c \bar{K}(20\%)$ \\

                &$0(3/2^-)$ &$(0.68,6.24)$  &$\Xi_c\bar{K}(66\%)$ &$\Xi_c^* \bar{K}(33\%)$    \\
                
                 &$1(3/2^-)$   &$(0.21,1.00)$    &$\Xi_c^* \bar{K}(75\%)$ &$\Xi_c\bar{K}^*(23\%)$ \\
               
\bottomrule[1pt]\bottomrule[1pt]
\end{tabular}
\end{table*}

The ongoing discovery of new hadronic states presents the crucial scientific challenge of distinguishing conventional hadrons from exotic configurations. In this context, the study of strong decay properties emerges as a powerful and essential tool for elucidating the internal structure of these states.

In this work, we have systematically investigated the two-body strong decay behaviors of possible $Y_c\bar{K}^{(*)}$ molecular pentaquarks with $Y_c=(\Lambda_c, \Sigma_c, \Xi_c, \Xi_c^{\prime})$ using an effective Lagrangian approach. Within this framework, the decay proceeds via the scattering of the molecular constituents into two final-state hadrons through meson or baryon exchange. The specific molecular states studied encompass those with strangeness $|S|=1$, previously predicted in our earlier work \cite{Chen:2023qlx}, and new candidates with $|S|=2$, identified in the present study using the one-boson-exchange (OBE) model with consistent coupling constants. The full list includes:

\begin{itemize}
    \item $\Sigma_c\bar{K}$ molecule with $1/2(1/2^-)$.
    \item $\Sigma_c\bar{K}^*$ molecules with $1/2(1/2^-,3/2^-)$, and $3/2(3/2^-)$.
    \item $\Lambda_c\bar{K}^*/\Sigma_c\bar{K}^*$ coupled molecule with $1/2(1/2^-)$.
    \item $\Xi_c^{\prime}\bar{K}/\Xi_c\bar{K}^*/\Xi_c^{\prime}\bar{K}^*$ coupled molecule with $0(1/2^-)$.
    \item $\Xi_c\bar{K}^*/\Xi_c^{\prime}\bar{K}^*$ coupled molecules with $0(1/2^-, 3/2^-)$.
    \item $\Xi_c^{\prime}\bar{K}^*$ molecules with $0(1/2^-, 3/2^-)$ and $1(1/2^-)$.
\end{itemize}

A comprehensive summary of their two-body strong decay properties is presented in Table \ref{summary}. Our key findings are:

\begin{enumerate}
    \item The total decay widths for these molecular states range from several to several tens of MeV.
    \item The corresponding branching ratios are remarkably stable, showing minimal variation with binding energy.
    \item The decay patterns exhibit a clear preference for final states consisting of a charmed baryon and a strange meson. This preference is driven by the dominant role of light meson (particularly $\pi$ and $\rho$) exchange processes.
    \item Coupled-channel effects are shown to play an essential role, not only in the formation of several candidates but also in shaping their decay amplitudes.
\end{enumerate}

The predicted decay widths and distinctive branching ratios (summarized in Table \ref{summary}) provide a clear roadmap for experimental identification. Dedicated amplitude analyses of processes such as $\Lambda_b$ decays at LHCb or Belle II, with focus on the identified dominant channels are strongly encouraged. The experimental confirmation or exclusion will significantly advance our understanding of exotic hadron spectroscopy and offer profound insights into the low-energy dynamics of QCD in multiquark systems.

\section*{ACKNOWLEDGMENTS}\label{sec5}

This project is supported by the National Natural Science Foundation of China under Grants Nos. 12305139 and 12305087. Rui Chen is also supported by the Xiaoxiang Scholars Programme of Hunan Normal University.

\appendix\label{app}

\section{OBE effective potentials and mass spectrum for possible $\Xi_c^{(\prime)}\bar{K}^{(*)}$ molecules}

The effective Lagrangians for describing the interactions between the heavy baryons and light meson $(\sigma, \pi, \eta, \rho, \omega)$ are constructed in the heavy quark symmetry and chiral symmetry \cite{Liu:2011xc}. The relevant expressions read as
\begin{eqnarray}
\mathcal{L}_{\mathcal{B}_{\bar{3}}} &=& l_B\langle\bar{\mathcal{B}}_{\bar{3}}\sigma\mathcal{B}_{\bar{3}}\rangle
          +i\beta_B\langle\bar{\mathcal{B}}_{\bar{3}}v^{\mu}(\mathcal{V}_{\mu}-\rho_{\mu})\mathcal{B}_{\bar{3}}\rangle,\label{lag1}\\
\mathcal{L}_{\mathcal{B}_{6}} &=& 
i\beta_{S}\langle\bar{\mathcal{S}}_{\mu}v_{\alpha}
    \left(\mathcal{V}_{ab}^{\alpha}-\rho_{ab}^{\alpha}\right) \mathcal{S}^{\mu}\rangle
    +\lambda_S\langle\bar{\mathcal{S}}_{\mu}F^{\mu\nu}(\rho)\mathcal{S}_{\nu}\rangle\nonumber\\
    &&+l_S\langle\bar{\mathcal{S}}_{\mu}\sigma\mathcal{S}^{\mu}\rangle
         -\frac{3}{2}g_1\varepsilon^{\mu\nu\lambda\kappa}v_{\kappa} \langle\bar{\mathcal{S}}_{\mu}A_{\nu}\mathcal{S}_{\lambda}\rangle,
         \\
\mathcal{L}_{\mathcal{B}_{\bar{3}}\mathcal{B}_6} &=& ig_4\langle\bar{\mathcal{S}^{\mu}}A_{\mu}\mathcal{B}_{\bar{3}}\rangle+i\lambda_I\varepsilon^{\mu\nu\lambda\kappa}v_{\mu}\langle \bar{\mathcal{S}}_{\nu}F_{\lambda\kappa}\mathcal{B}_{\bar{3}}\rangle+h.c..\label{lag2}
\end{eqnarray}
Here, $A_{\mu}$ and $\mathcal{V}_{\mu}$ denote the axial and vector currents, respectively, i.e.,
\begin{eqnarray*}
A_{\mu} &=& \frac{1}{2}(\xi^{\dag}\partial_{\mu}\xi-\xi\partial_{\mu}\xi^{\dag})=\frac{i}{f_{\pi}}
\partial_{\mu}{P}+\ldots,\\
\mathcal{V}_{\mu} &=&
\frac{1}{2}(\xi^{\dag}\partial_{\mu}\xi+\xi\partial_{\mu}\xi^{\dag})
=\frac{i}{2f_{\pi}^2}\left[{P},\partial_{\mu}{P}\right]+\ldots,
\end{eqnarray*}
with $\xi=\text{exp}(i{P}/f_{\pi})$ and $f_{\pi}=132$ MeV. And $v=(1,\textbf{0})$ is the four velocity, $\rho_{ba}^{\mu}=ig_V{V}_{ba}^{\mu}/\sqrt{2}$, and $F^{\mu\nu}(\rho)=\partial^{\mu}\rho^{\nu}-\partial^{\nu}\rho^{\mu}
+\left[\rho^{\mu},\rho^{\nu}\right]$. $\mathcal{B}_{\bar{3}}$ and $\mathcal{S}_{\mu} =-\sqrt{\frac{1}{3}}(\gamma_{\mu}+v_{\mu})\gamma^5\mathcal{B}_6+\mathcal{B}_{6\mu}^*$ denote the ground heavy baryons multiplets with their light quarks in the $\bar{3}$ and $6$ flavor representation, respectively. The matrices $\mathcal{B}_{\bar{3}}$, $\mathcal{B}_6$, ${P}$, and ${V}$ read as 
\begin{eqnarray}
\mathcal{B}_{\bar{3}} = \left(\begin{array}{ccc}
0               &\Lambda_c^+  & \Xi_{c}^{+}  \\
-\Lambda_c^+    &0            & \Xi_{c}^{0} \\
-\Xi_{c}^{+}    &-\Xi_{c}^{0}           & 0 \\
\end{array}\right),\quad
\mathcal{B}_6 = \left(\begin{array}{ccc}
\Sigma_c^{++}    &\frac{\Sigma_c^{+}}{\sqrt{2}} &\frac{\Xi_{c}^{\prime+}}{\sqrt{2}}\\
\frac{\Sigma_c^{+}}{\sqrt{2}}   &\Sigma_c^{0}
&\frac{\Xi_{c}^{\prime0}}{\sqrt{2}} \\
\frac{\Xi_{c}^{\prime+}}{\sqrt{2}}   &\frac{\Xi_{c}^{\prime0}}{\sqrt{2}}&\Omega_{c}^{0} \\
\end{array}\right).
\end{eqnarray}
Here, we also take the same parameters in Refs.\cite{Belle:2004zjl}, where $g_1=1$ are extracted from the experimental width of $\Sigma_c\to\Lambda_c+\pi$ . The remaining coupling constants are estimated from the nucleon-nucleon interactions with the help of the quark model \cite{Liu:2011xc,Chen:2017xat}, $l_S=-2l_B=7.3$, $g_1=(\sqrt{8}/3)g_4=1.0$, $\beta_Sg_V=-2\beta_Bg_V=12.0$, $\lambda_Sg_V=-2\sqrt{2}\lambda_Ig_V=19.2~ \text{GeV}^{-1}$.

The flavor wave functions for the $\Xi_c^{(\prime)}\bar{K}^{(*)}$ systems can be expressed as
\begin{eqnarray}
 |0,0\rangle &=&\frac{1}{\sqrt{2}}
 \left(|\Xi_c^{(\prime,*)+}\bar{K}^{(*)-}\rangle +\Xi_c^{(\prime,*)0}\bar{K}^{(*)0}\rangle\right),\\
 |1,1\rangle &=&-\left|\Xi_c^{(\prime,*)+}\bar{K}^{(*)0} \rangle\right.,\\
 |1,0\rangle &=&\frac{1}{\sqrt{2}}
 \left(|\Xi_c^{(\prime,*)+}\bar{K}^{(*)-} \rangle - |\Xi_c^{(\prime,*)0}\bar{K}^{(*)0}\rangle\right),\\
|1,-1\rangle &=&\left|\Xi_c^{(\prime,*)0}\bar{K}^{(*)-} \rangle\right.,
\end{eqnarray}
After considering the $S-D$ wave mixing effects, the spin-orbit wave functions for the involved systems can be constructed as
 \begin{eqnarray}
 \nonumber \Xi_c^{(\prime)}\bar{K}     && (J^P=1/2^-): \left|^2S_{1/2}\right\rangle,\\ \nonumber
\Xi_c^{(\prime)}\bar{K}^{*} &&(J^P=1/2^-): \left|^2S_{1/2}\right\rangle,\left|^4D_{1/2}\right\rangle,\\
\Xi_c^{(\prime)}\bar{K}^{*} &&(J^P=3/2^-): \left|^4S_{3/2}\right\rangle,\left|^2D_{3/2}\right\rangle,\left|^4D_{3/2}\right\rangle.
\end{eqnarray}

\begin{table*}[!htbp]
\centering
\caption{Here, $A=l_{B}g_{\sigma},B=\beta_{s}g_{v},C=\lambda_I g_V,D=\frac{g_{4}}{f_{\pi}},E=l_{S}g_{\sigma},
F=\lambda_S g_V,H=\frac{g_{1}}{f_{\pi}}$. $Y(\Lambda,m,r) = \frac{1}{4\pi r}(e^{-mr}-e^{-\Lambda r})-\frac{\Lambda^2-m^2}{8\pi\Lambda} e^{-\Lambda r}$, $\mathcal{Y}_{\Lambda,m_{a}}^{ij} = \mathcal{D}_{ij}Y(\Lambda,m_{a},r)$, $\mathcal{X}_{\Lambda,m_{a}}^{ij} = \mathcal{D}_{ij}Z(\Lambda,m_{a},r)$, $\mathcal{Z}_{\Lambda,m_{a}}^{ij} = \left(\mathcal{E}_{ij}\nabla^2+\mathcal{F}_{ij}r\frac{\partial}{\partial r }\frac{1}{r}\frac{\partial}{\partial r}\right)Y(\Lambda,m_{a},r)$, and $\mathcal{Z}_{\Lambda,m_{a}}^{\prime ij} = \left(2\mathcal{E}_{ij}\nabla^2-\mathcal{F}_{ij}r\frac{\partial}{\partial r }\frac{1}{r}\frac{\partial}{\partial r}\right)Y(\Lambda,m_{a},r)$.}
\label{obe}
\renewcommand\tabcolsep{0.15cm}
\renewcommand{\arraystretch}{2.2}
\begin{tabular}{c|l}
\toprule[1pt]\toprule[1pt]
\textbf{Processes}   &\textbf{OBE effective potentials}\\
 \hline
$\Xi_{c}\bar{K} \rightarrow \Xi_c \bar{K}$
&$-AY(\Lambda,m_\sigma,r)-\frac{B g_{PPV}}{2\sqrt{2}}\left(\mathcal{G}Y(\Lambda,m_\rho,r)
 +Y(\Lambda,m_\omega,r)
 -2Y(\Lambda,m_\phi,r)\right)$ \\
 $\Xi_c\bar{K} \rightarrow \Xi_c^{\prime}\bar{K}$  &
$-\frac{ C g_{PPV}}{2\sqrt{3}M_{\bar{K}}}\left(\mathcal{G} W(\Lambda,m_{\rho12},r)+W(\Lambda, m_{\omega12},r)+2W(\Lambda, m_{\phi12}, r)\right)$\\
$\Xi_c\bar{K} \rightarrow \Xi_c^{\prime} \bar{K}^{*}$& $\frac{D g_{PPV}}{16\sqrt{6}\sqrt{M_{\bar{K}} M_{\bar{K^*}}}}  \left(1-\frac{q_{14}}{2M_{\Xi_c^{\prime}}}\right)\left(\mathcal{G}\mathcal{Z}_{\Lambda_{14},m_{\pi14}}^{14}+\mathcal{Z}_{\Lambda_{14},m_{\eta14}}^{14}\right)-\frac{C g_{VVP}}{6\sqrt{6}\sqrt{M_{\bar{K}}}} \left(1-\frac{q_{14}}{M_{\bar{K^*}}}\right)\left(\frac{\mathcal{G}}{m_{\rho_{14}}}\mathcal{Z}_{\Lambda_{14},m_{\rho14}}^{\prime14}
+\frac{1}{m_{\omega_{14}}}\mathcal{Z}_{\Lambda_{14},m_{\omega14}}^{\prime14}+\frac{2}{m_{\phi_{14}}}\mathcal{Z}_{\Lambda_{14},m_{\phi14}}^{\prime14}\right)$\\
$\Xi_{c}^{\prime}\bar{K} \rightarrow \Xi_c^{\prime} \bar{K}$&
$-\frac{1}{2}EY(\Lambda,m_\sigma,r)
 +\frac{B g_{PPV}}{4\sqrt{2}}
 \left(\mathcal{G}Y(\Lambda,m_\rho,r)
 +Y(\Lambda,m_\omega,r)
 -2Y(\Lambda,m_\phi,r)\right)+\frac{F g_{PPV}}{12\sqrt{2}M_{\Xi_{c}^{\prime}}}
 \left(\mathcal{G}Z(\Lambda,m_\rho,r)
 +Z(\Lambda,m_\omega,r)
 -2Z(\Lambda,m_\phi,r)\right)$\\
 
 $\Xi_c\bar{K}^{*} \rightarrow \Xi_c\bar{K}^{*}$&
 $-A\mathcal{Y}_{\Lambda,m_{\sigma}}^{24}+\frac{B g_{VVV}}{2\sqrt{2}} \left(\mathcal{G}\mathcal{Y}_{\Lambda,m_{\rho}}^{24}+\mathcal{Y}_{\Lambda,m_{\omega}}^{24} - 2\mathcal{Y}_{\Lambda,m_{\phi}}^{24}\right)$\\
 
$\Xi_c^{\prime}\bar{K} \rightarrow \Xi_c^{\prime} \bar{K}^{*}$&
$-\frac{3Hg_{PPV}}{16\sqrt{M_{\bar{K}} M_{\bar{K^*}}}}
\left(\mathcal{G}\mathcal{Z}_{\Lambda_{24},m_{\pi24}}^{24}+\frac{1}{3}\mathcal{Z}_{\Lambda_{24},m_{\eta24}}^{24}\right)-\frac{9Fg_{VVP}}{\sqrt{2}\sqrt{M_{\bar{K}}M_{\bar{K}^{*}}}}\left(\frac{1}{4M_{\Xi_{c}^{\prime}}}-\frac{\sqrt{M_{\bar{K^*}}^3}}{\sqrt{M_{\bar{K}}}}+\frac{q_{24}}{2}\right)\left(\mathcal{G}\mathcal{Z}_{\Lambda_{24},m_{\rho24}}^{\prime24}
+\mathcal{Z}_{\Lambda_{24},m_{\omega24}}^{\prime24}-2\mathcal{Z}_{\Lambda_{24},m_{\phi24}}^{\prime24}\right)$\\

$\Xi_{c}\bar{K}^{*} \rightarrow \Xi_{c}^{\prime}\bar{K}^{*}$
&$\frac{D g_{VVP}}{4\sqrt{3}\sqrt{m_{\bar{K}^{*}}}}
\left(\mathcal{G}\mathcal{Z}_{\Lambda_{34},m_{\pi34}}^{34}+\mathcal{Z}_{\Lambda_{34},m_{\eta14}}^{34}\right)+\frac{C g_{VVV}}{4\sqrt{6}M_{K^*}}
\left(\mathcal{G}\mathcal{Z}_{\Lambda_{34},m_{\rho34}}^{\prime34}
+\mathcal{Z}_{\Lambda_{34},m_{\omega34}}^{\prime34}+2\mathcal{Z}_{\Lambda_{34},m_{\phi34}}^{\prime34}\right)$\\

$\Xi_{c}^{\prime}\bar{K}^{*} \rightarrow \Xi_{c}^{\prime}\bar{K}^{*}$&$-\frac{1}{2}E\mathcal{Y}_{\Lambda,m_{\sigma}}^{44}+ \frac{\sqrt{3}Hg_{VVP}}{8\sqrt{2}} \left(\mathcal{G}\mathcal{Z}_{\Lambda_{44},m_{\pi34}}^{44}+\frac{1}{3}\mathcal{Z}_{\Lambda_{44},m_{\eta44}}^{44}\right)- \frac{B g_{VVV}}{4\sqrt{2}} \left(\mathcal{G}\mathcal{Y}_{\Lambda,m_{\rho}}^{44} +\mathcal{Y}_{\Lambda,m_{\omega}}^{44} - 2\mathcal{Y}_{\Lambda,m_{\phi}}^{44} \right)$\\
&$-\frac{F g_{VVV}}{12\sqrt{2}M_{\Xi_c^*}} 
\left(\mathcal{G}\mathcal{X}_{\Lambda,m_{\rho}}^{44} +\mathcal{X}_{\Lambda,m_{\omega}}^{44} - 2\mathcal{X}_{\Lambda,m_{\phi}}^{44} \right)- \frac{F g_{VVV}}{6\sqrt{2}M_{\bar{K}^*}}\left(\mathcal{G}\mathcal{Z}_{\Lambda_{44},m_{\rho44}}^{\prime44}
+\mathcal{Z}_{\Lambda_{44},m_{\omega44}}^{\prime44}-2\mathcal{Z}_{\Lambda_{44},m_{\phi44}}^{\prime44}\right)$
 \\\hline
\bottomrule[1pt]\bottomrule[1pt]

\end{tabular}
\end{table*}

Using the same procedures and effective Lagragians in our previous work , we can deduce the OBE effective potentials for the $\Xi_c^{(\prime)}\bar{K^{(*)}}$ systems, as collected in Table \ref{obe}.

\begin{figure*}[!htbp]
    \centering
        \includegraphics[width=0.24\linewidth]{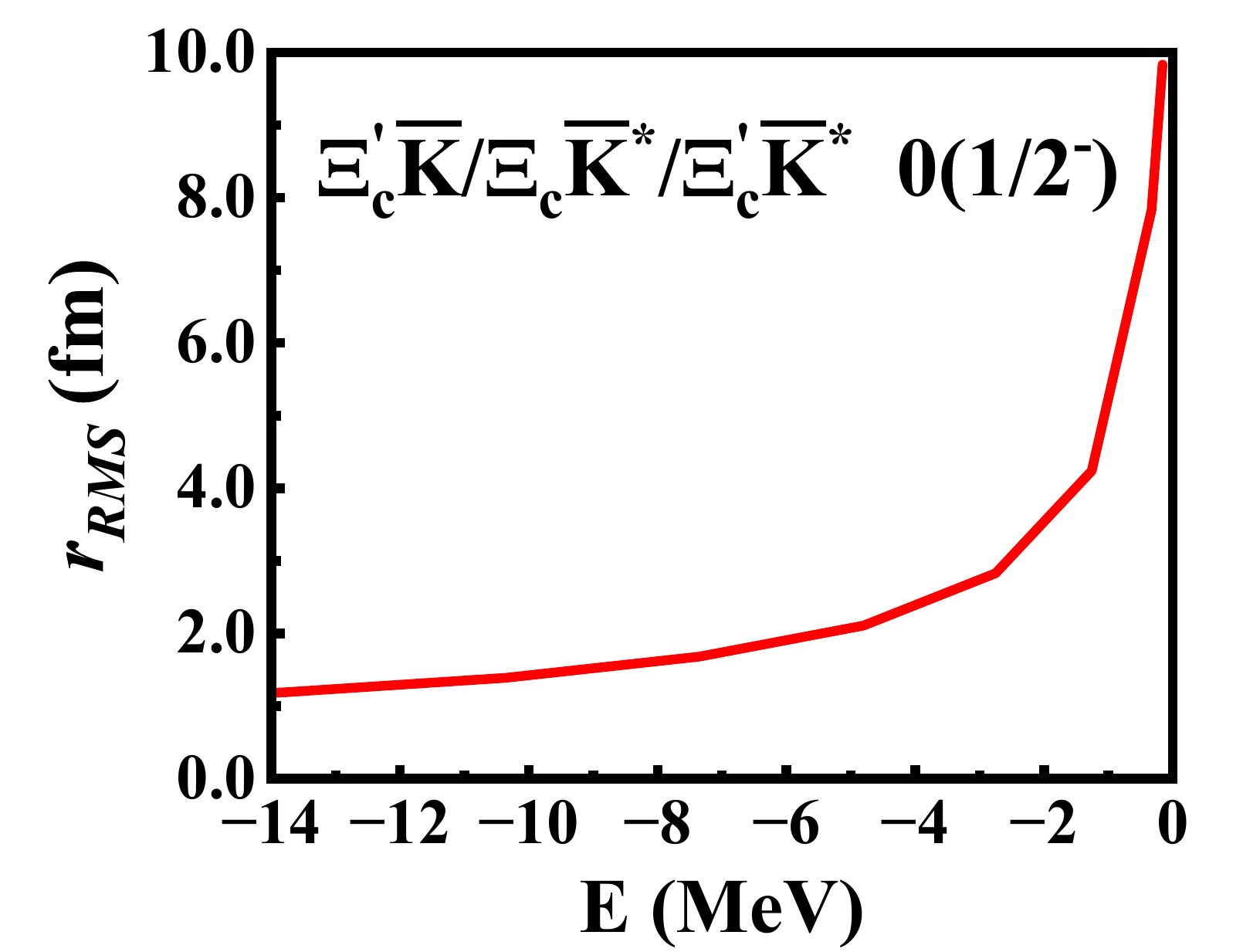}
        \includegraphics[width=0.24\linewidth]{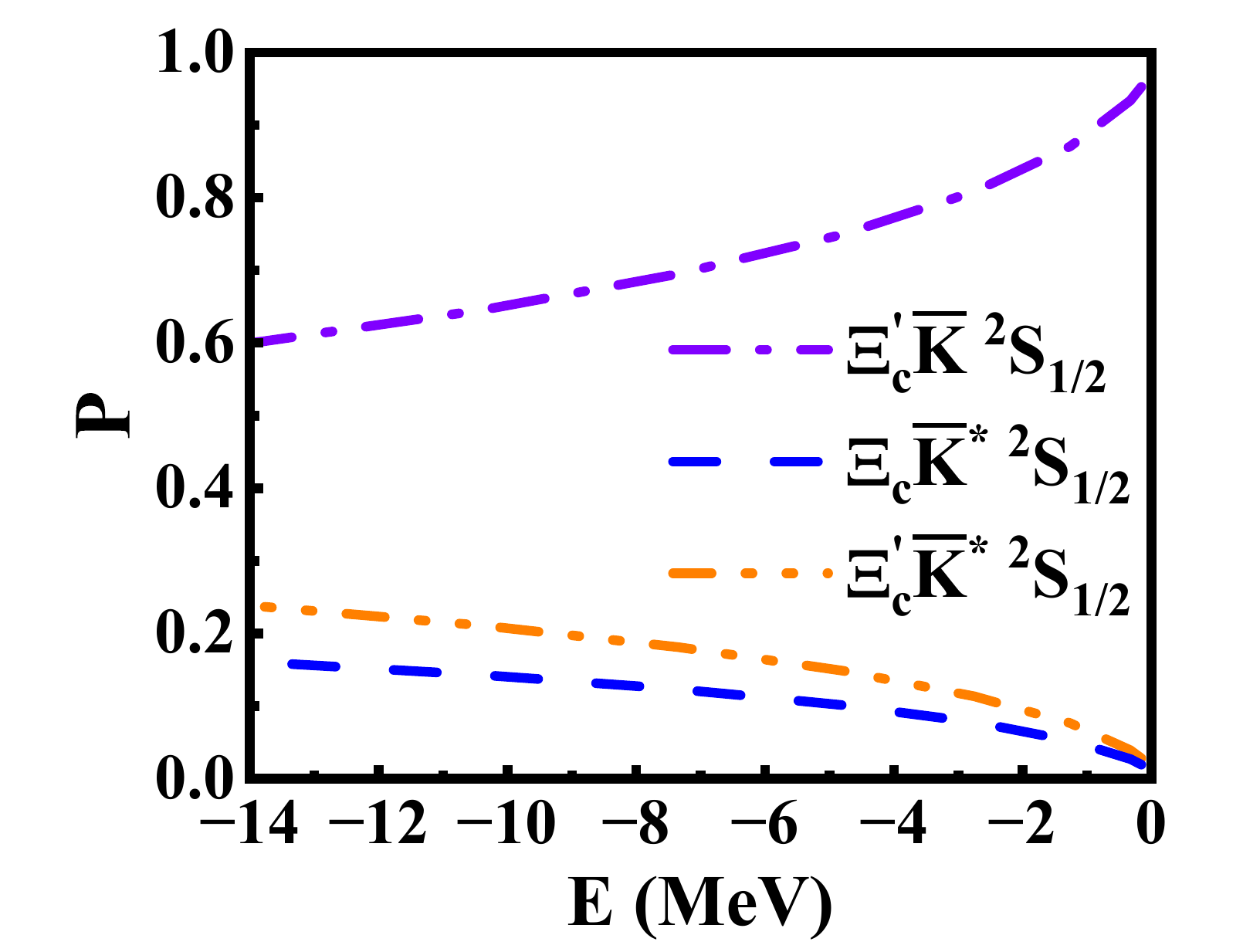}
         \includegraphics[width=0.24\linewidth]{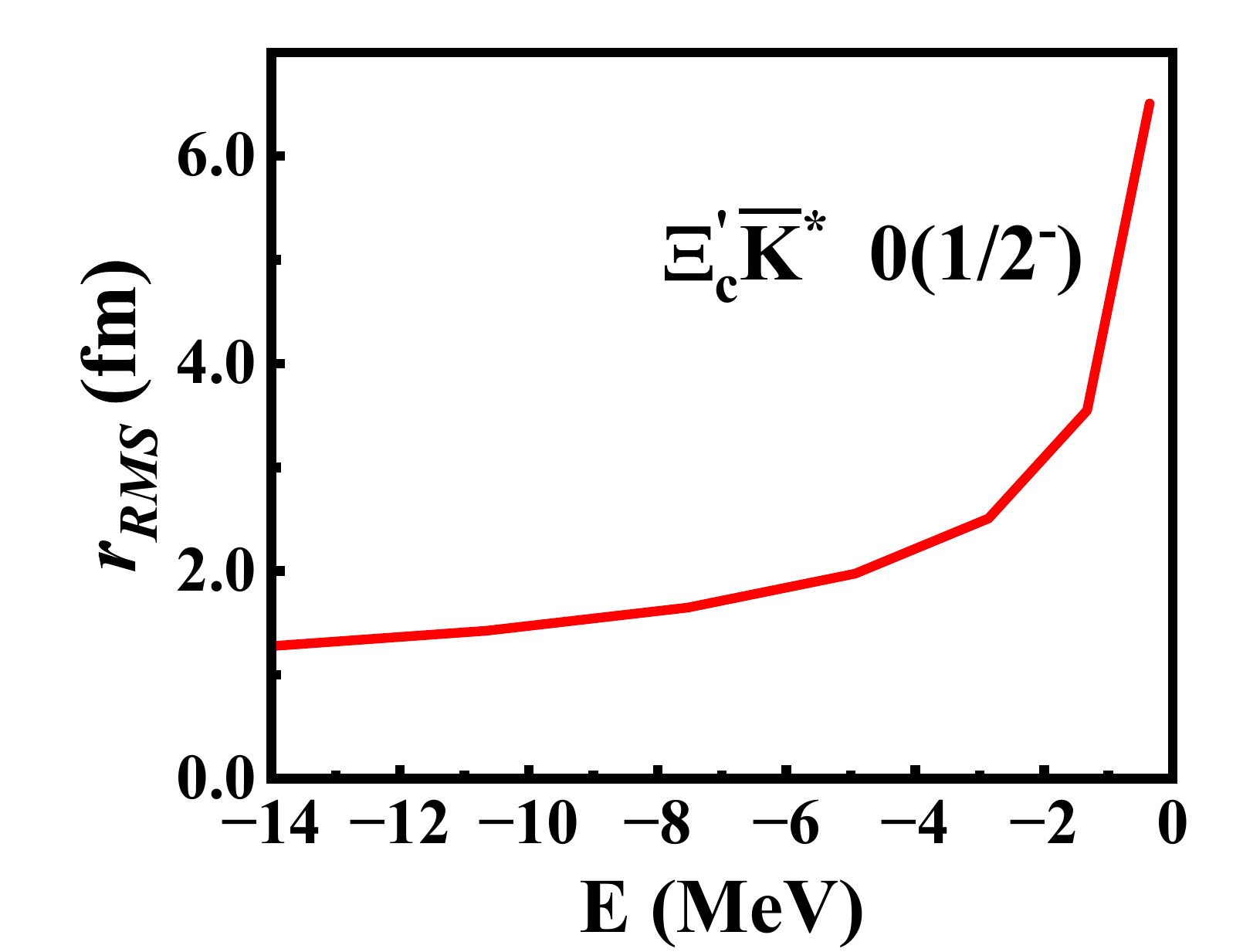}
        \includegraphics[width=0.24\linewidth]{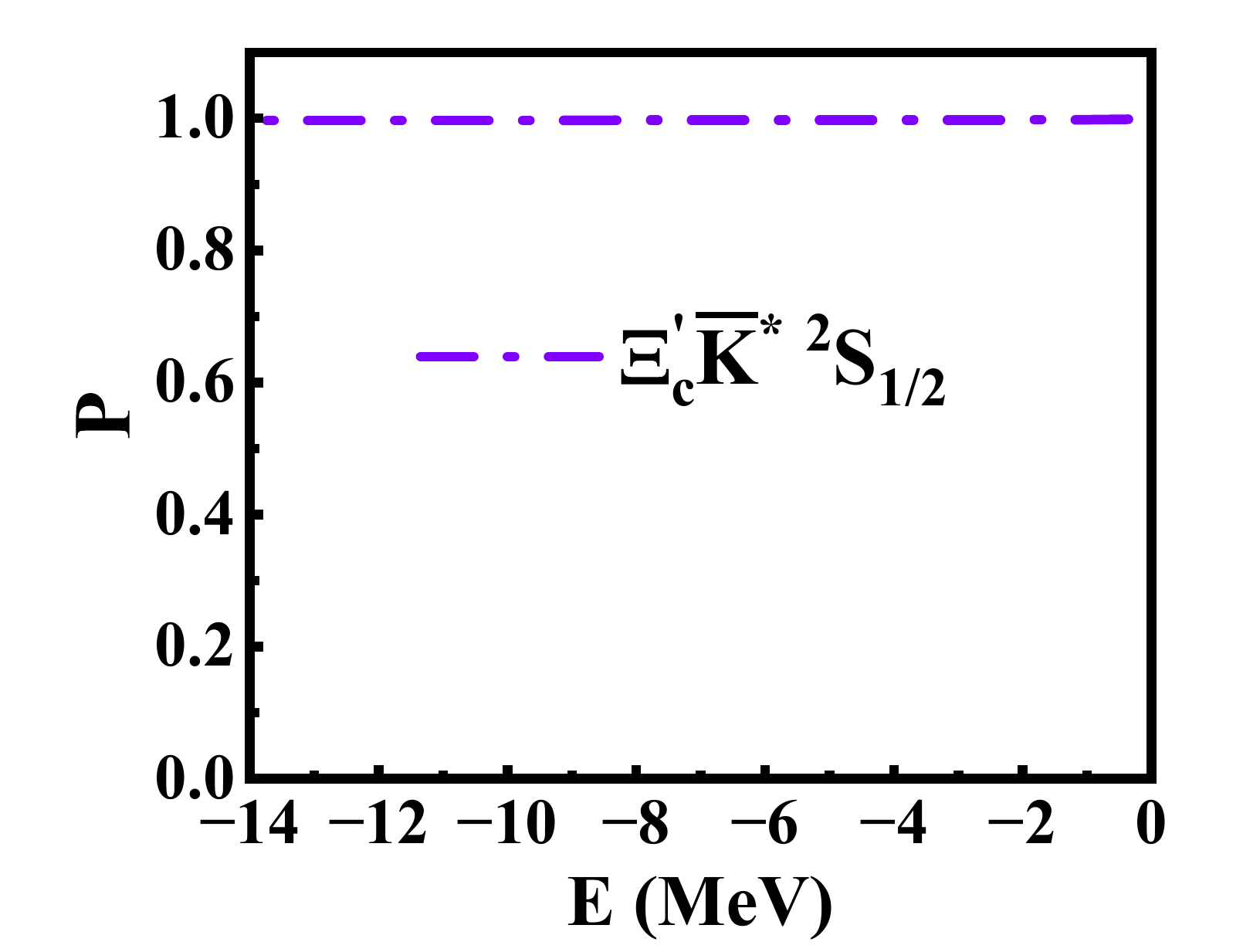}\\
        \includegraphics[width=0.24\linewidth]{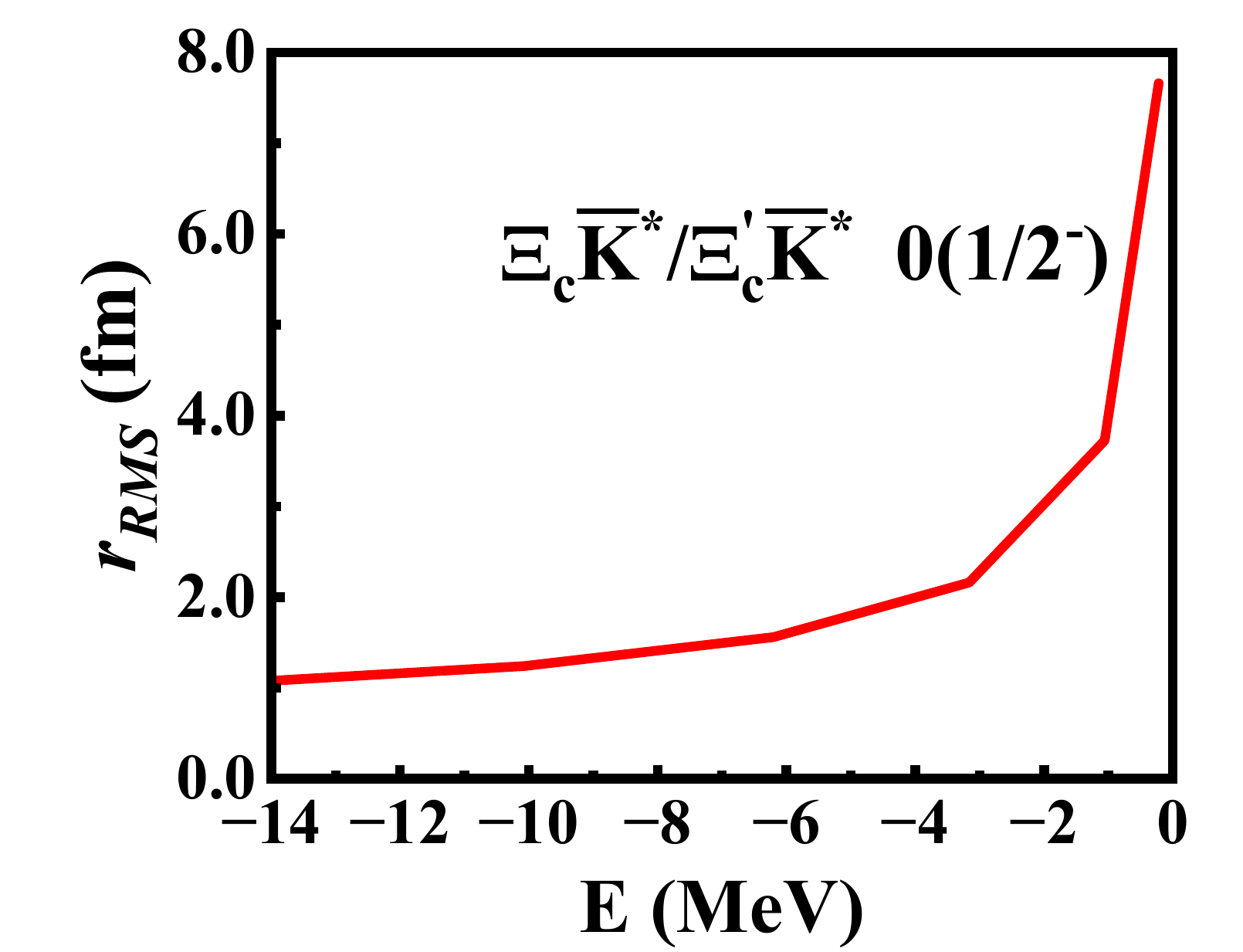}
        \includegraphics[width=0.24\linewidth]{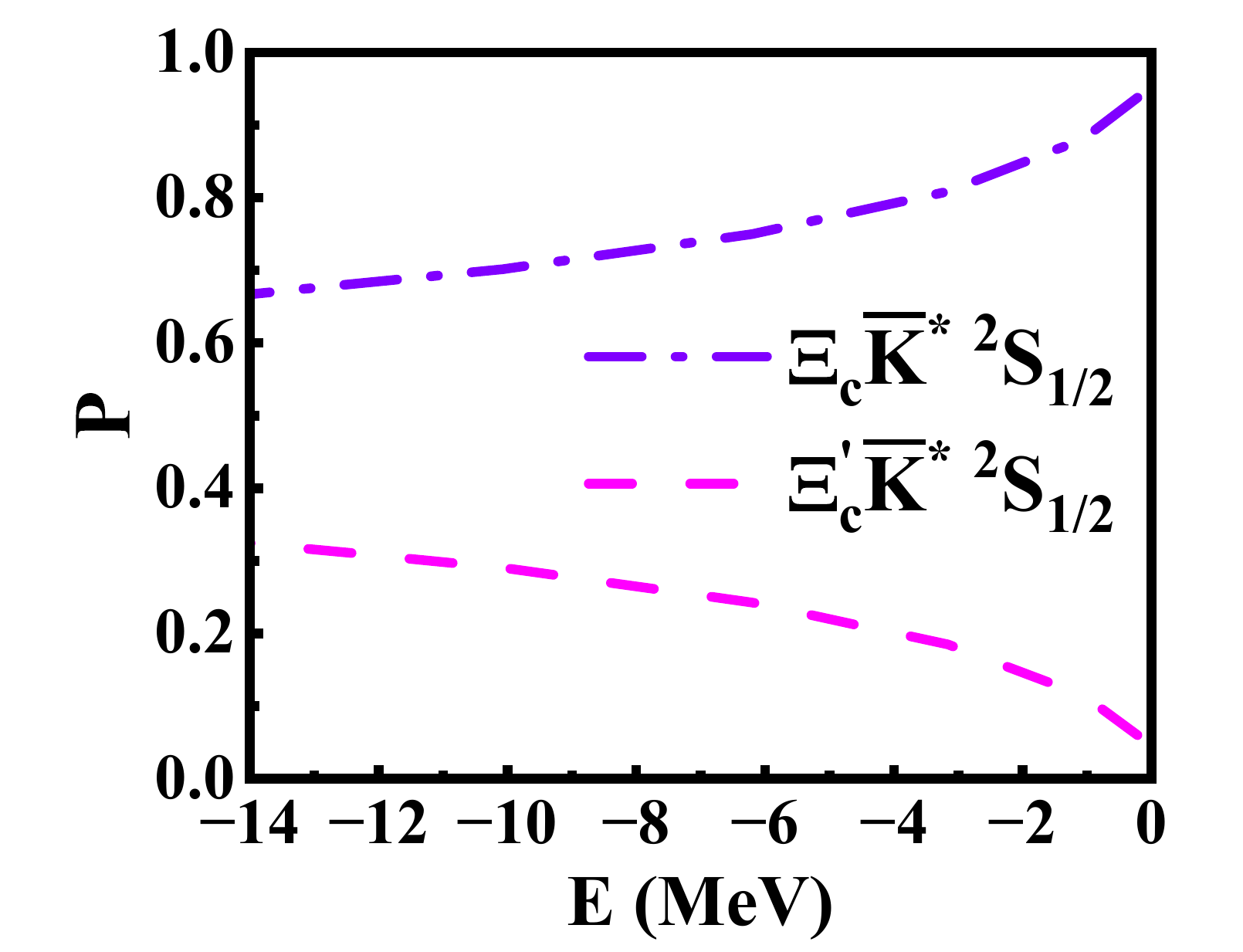}
        \includegraphics[width=0.24\linewidth]{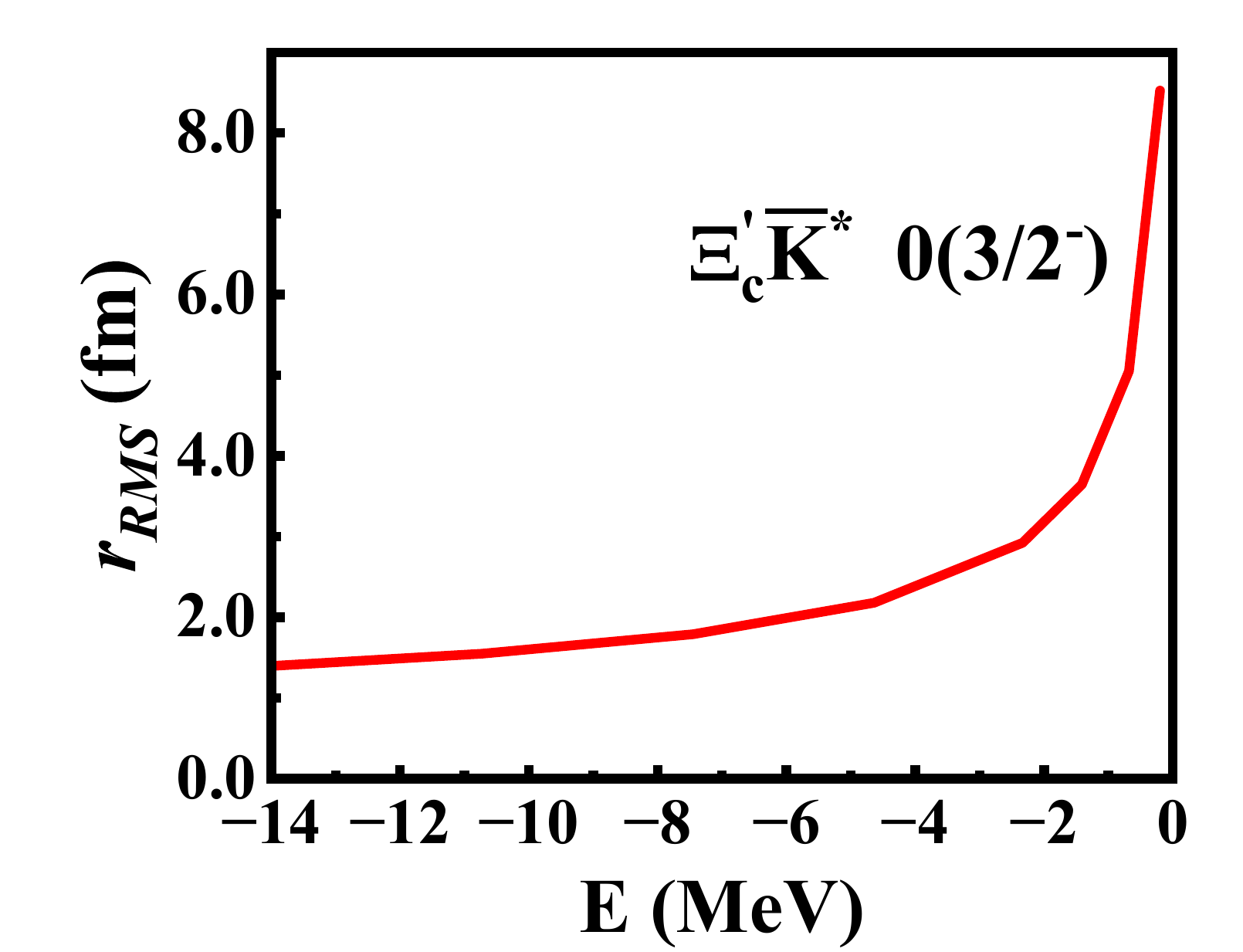}
       \includegraphics[width=0.24\linewidth]{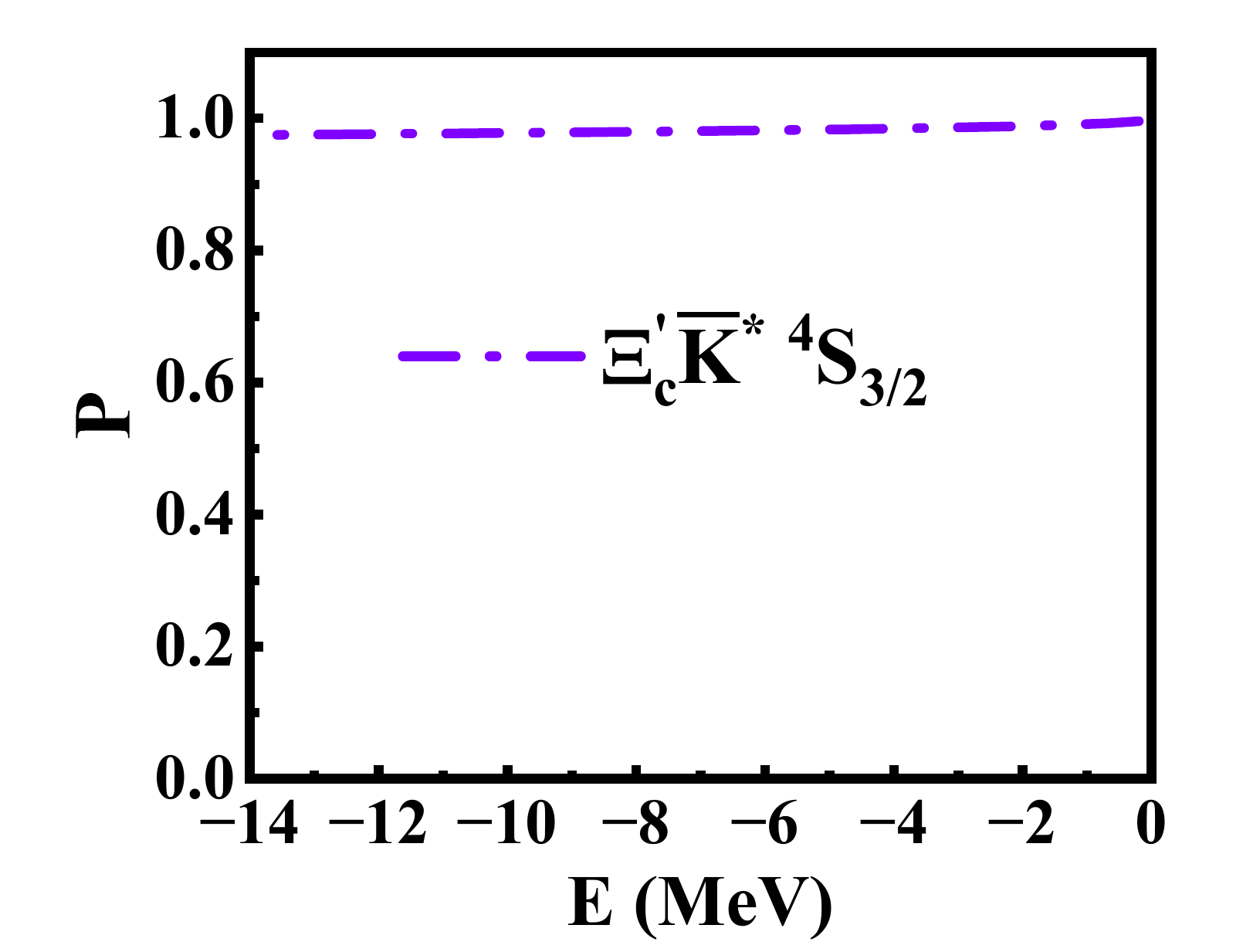}\\
       \includegraphics[width=0.24\linewidth]{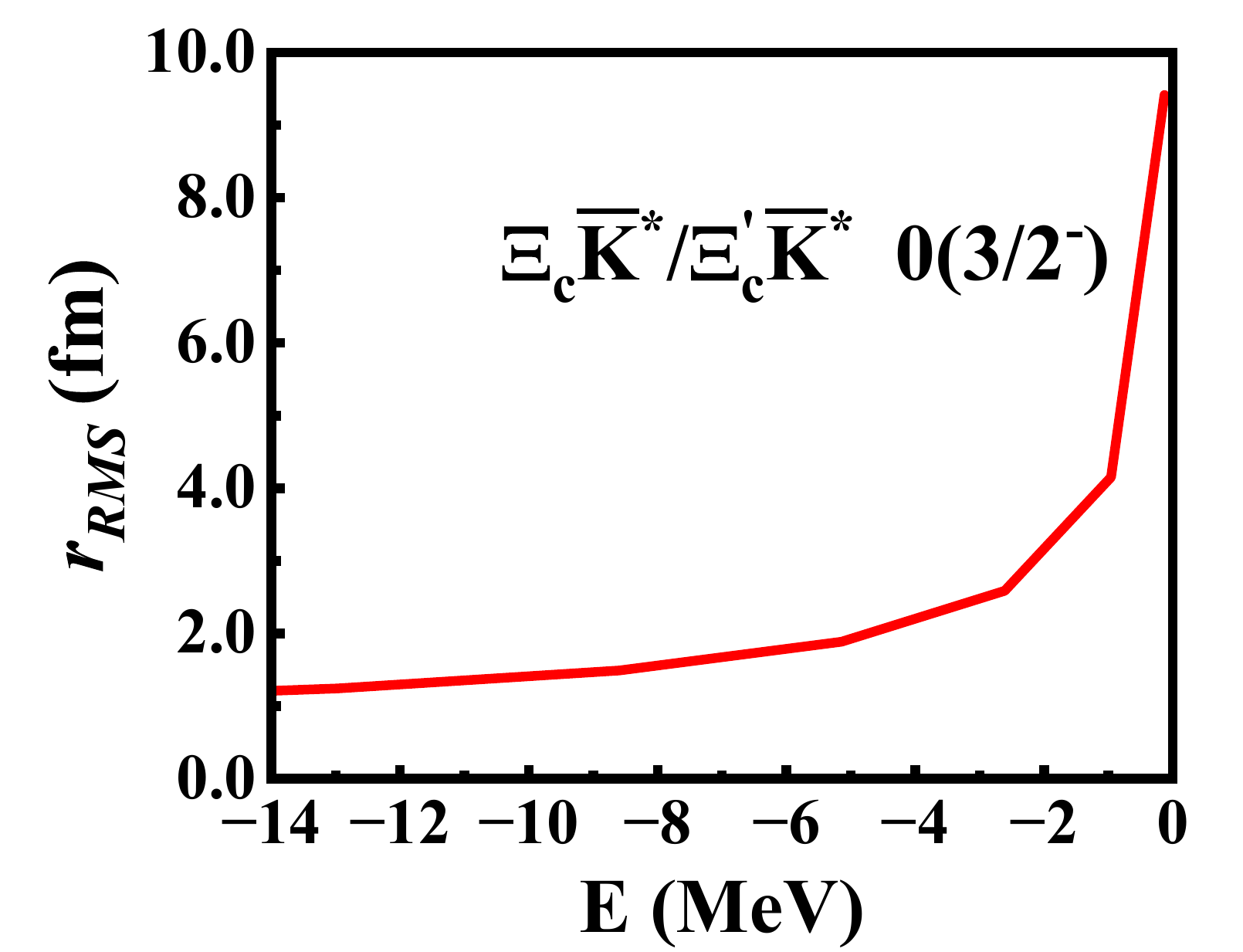}
        \includegraphics[width=0.24\linewidth]{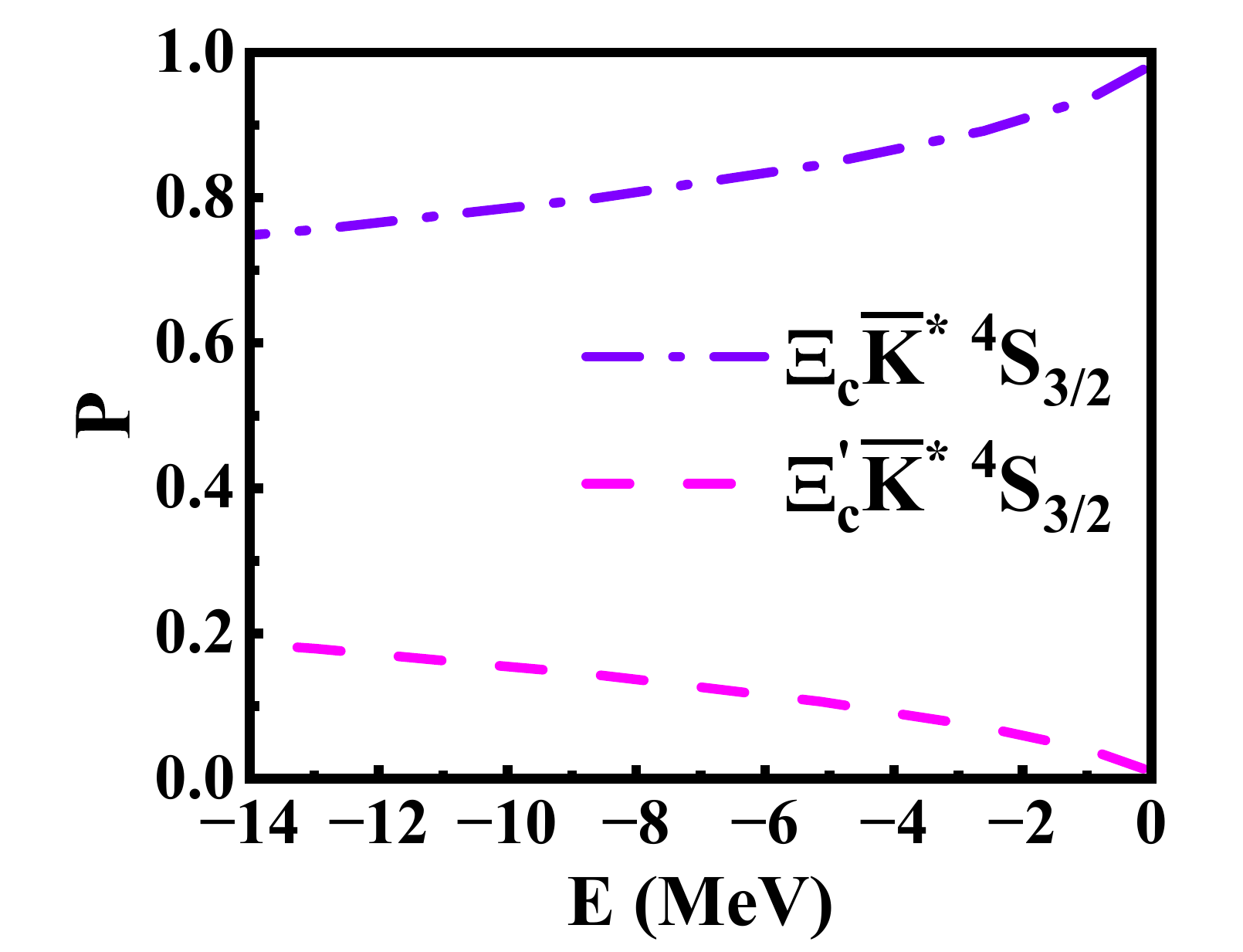}
        \includegraphics[width=0.24\linewidth]{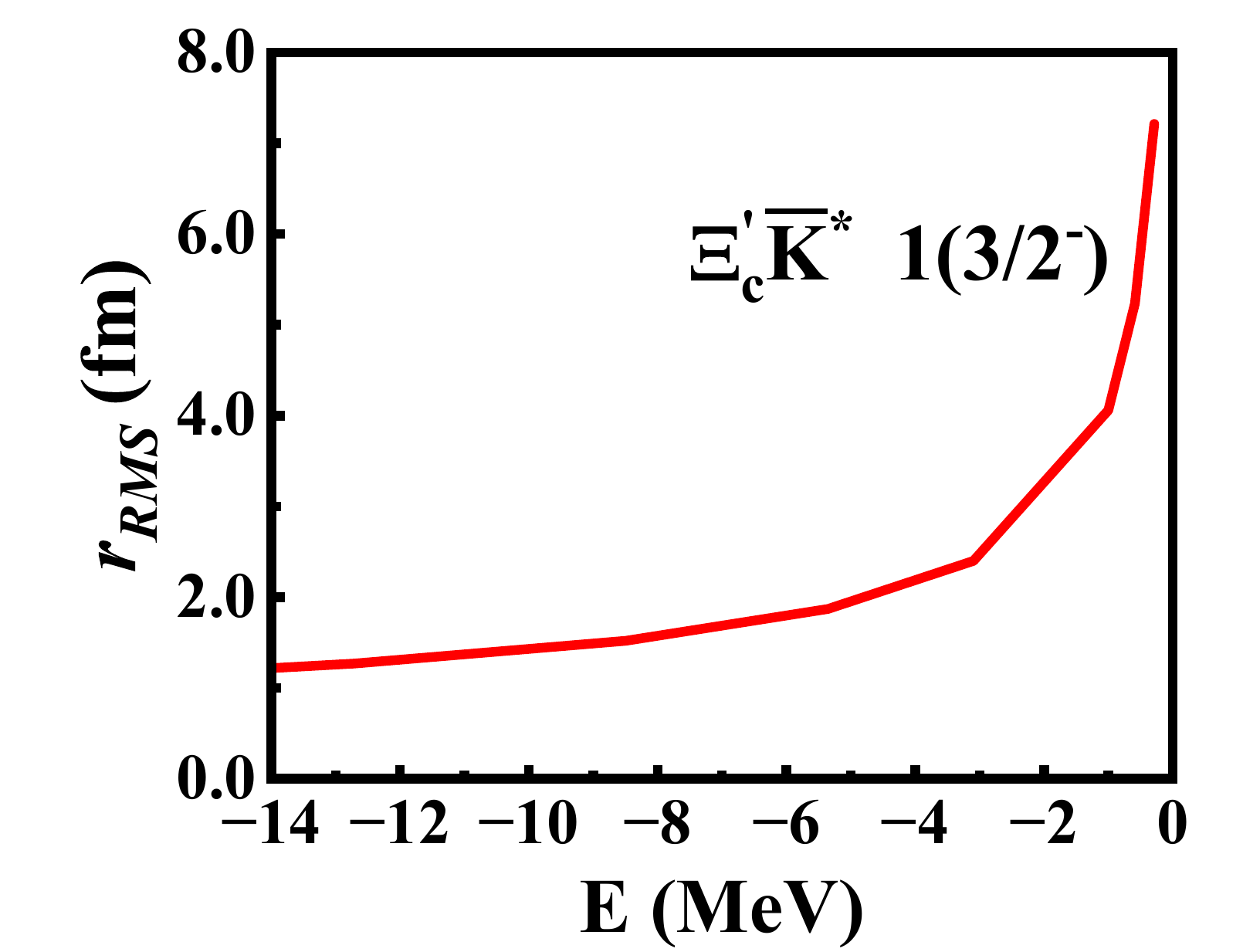}
        \includegraphics[width=0.24\linewidth]{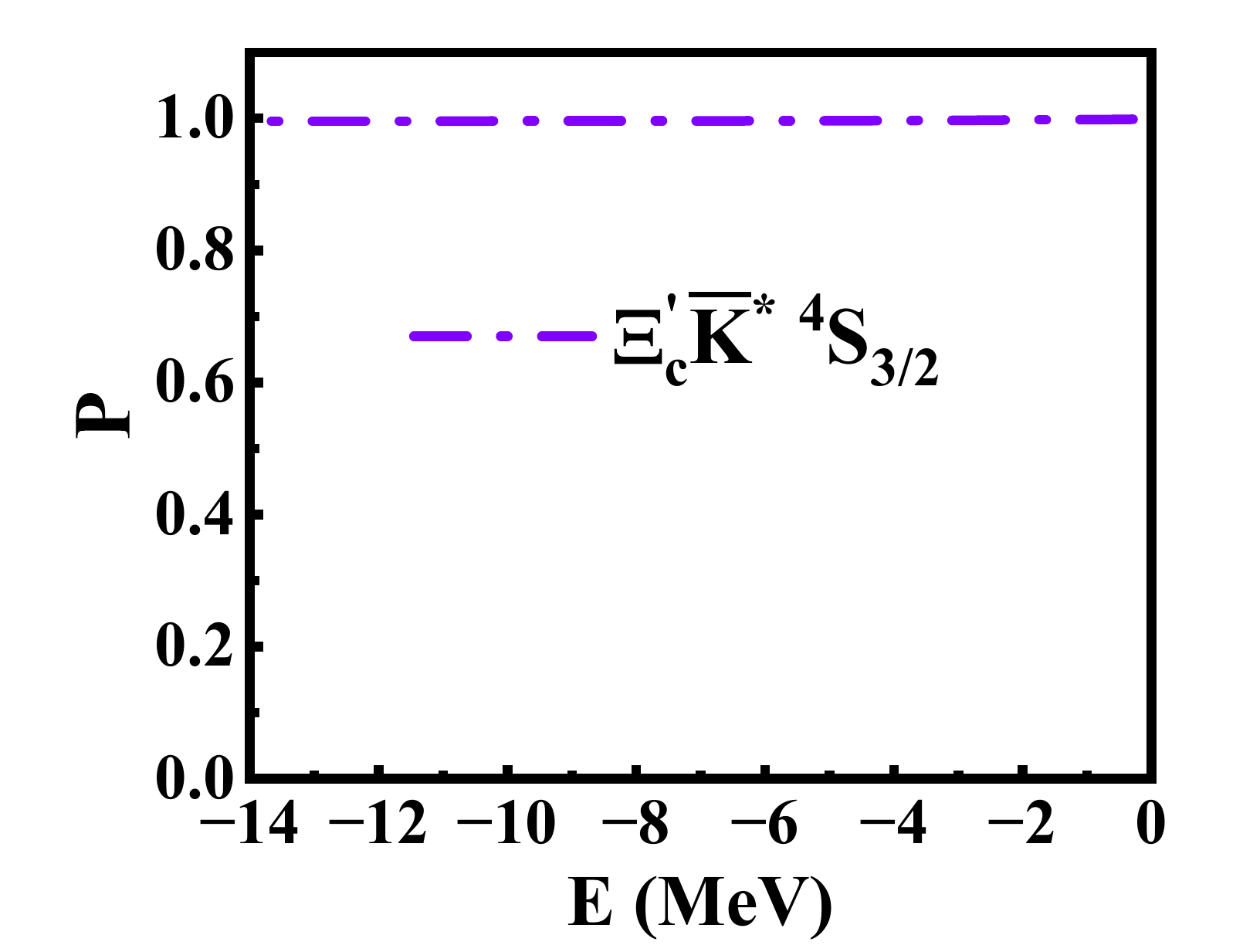}
     \caption{The bound state solutions (the binding energy $E$, the root-mean-square radius $r_{rms}$, and the probabilities $P$) for all the discussed channels of  the coupled ${\Xi^\prime_c} \bar{K}/{\Xi_c}\bar{K}^* /{\Xi^\prime_c}\bar{K}^*$ systems with $I(J^P) = 0(1/2^-)$, ${\Xi_c}\bar{K}^* /{\Xi^\prime_c}\bar{K}^*$ systems with $I(J^P) = 0(1/2^-)$ and $0(3/2^-)$, ${\Xi^\prime_c}\bar{K}^*$ systems with $I(J^P) = 0(1/2^-)$ , $0(3/2^-)$ and $1(3/2^-)$. }
      \label{num3}
\end{figure*}

After solving the coupled channel Schr\"{o}dinger equations  using the Gaussian Expansion Method to compute the mass spectrum, we can predict possible $\Xi_c^{(\prime)}\bar{K}^{(*)}$ molecular candidates, where their binding energies $E$ vary several to several tens MeV and the root-mean-square radius $r_{\text{RMS}}$ are around or larger than 1.00 fm. In Figure \ref{num3} we show plots of the root-mean-square radius $E$ and the $S$-wave branching ratio of the system with the binding energy $E$. In summary, our calculations suggest the following systems as possible molecular candidates: The $0(1/2^-)$ state in the coupled $\Xi_{c}^{\prime}\bar{K}$/$\Xi_c \bar{K}^{*}$/$\Xi_{c}^{\prime} \bar{K}^{*}$ channels, and our results  show that the coupled channel effects play an important role in the formative to the system;
The $0(1/2^-)$ and $0(3/2^-)$  states in the $\Xi_c\bar{K}^*$/$\Xi_c^{\prime}\bar{K}^*$ systems, where coupled-channel effects also are found to be insignificant; The $0(1/2^-)$, $0(3/2^-)$ and $1(3/2^-)$ states in the single-channel $\Xi_{c}^{\prime}\bar{K}^*$ system.

\section{Coupling constants and scattering amplitudes}

\begin{table*}[ht]
\renewcommand\tabcolsep{0.25cm}
\renewcommand{\arraystretch}{1.8}
\centering
\caption{The coupling constants adopted in our calculations. Here, we take $g_{NN\rho}=3.25$, $f_{NN\rho}=19.82$, $g_{NN\pi}=0.989$, $g_{\Delta N\pi}=2.13$, $g_{\Delta N\rho}=16.03$, $g_{VVP}=-7.07$, $g_{PPV}=3.02$, $g_{VVV}=2.30 $, and $\alpha_{BBV}=1.15$ \cite{Wang:2022oof}. } \label{coupling}
\small
\begin{tabular}{lcccccc}
\toprule
& \textbf{Coupling constant} &\textbf{Coefficient}
&\textbf{Coupling constant} &\textbf{Coefficient} &\textbf{Coupling constant} &\textbf{Coefficient}\\
\toprule[1pt]\toprule[1pt]
\midrule

$g_{NN\pi}$ & $g_{\Xi_{c}^{\prime}\Xi_{c}^{\prime}\eta}$ &$\frac{4}{5\sqrt{2}}$   &$g_{\Xi_{c}^{\prime}\Xi_{c}\pi}$ & $-\frac{\sqrt{6}}{5}$ &$g_{\Xi_c\Sigma_c \bar{K}}$ & $\frac{\sqrt{6}}{5}$\\
&$g_{\Xi_c^{\prime}\Sigma_c \bar{K}}$&$\frac{4}{5\sqrt{2}}$ &$g_{\Xi_c\Omega_c K}$&$-\frac{2\sqrt{3}}{5}$&$g_{\Xi_c^{\prime}\Omega_c K}$ & $\frac{4}{5}$\\
&$g_{\Xi_{c}\Xi D_s}$  &$-\frac{3\sqrt{3}}{5}$&$g_{\Xi_{c}^{\prime}\Xi D_s}$ &$\frac{1}{5}$&$g_{\Xi_{c}\Sigma D}$ & $ \frac{3\sqrt{3}}{5\sqrt{2}}$\\
&$g_{\Xi_{c}\Lambda D}$   & $ \frac{3}{5\sqrt{2}}$&$g_{\Xi_{c}^{\prime}\Sigma D}$     & $ \frac{1}{5\sqrt{2}}$&$g_{\Xi_{c}^{\prime}\Lambda D}$   & $-\frac{\sqrt{3}}{5\sqrt{2}} $\\
&$g_{\Sigma_c\Sigma_c\pi}$ &$\frac{4}{5}$ &$g_{\Sigma_c\Sigma_c\eta}$&$\frac{4}{5\sqrt{3}}$  &$g_{\Lambda_c\Sigma_c\pi}$&$\frac{2\sqrt{3}}{5}$   \\
&$g_{\Sigma_cDN}$&$\frac{1}{5}$  &$g_{\Lambda_c\Xi_c^{\prime} \bar{K}}$ & $\frac{\sqrt{6}}{5}$ &$g_{\Lambda_c N D}$&$-\frac{3\sqrt{3}}{5}$\\
&$g_{\Lambda_c \Lambda D_s}$&$\frac{3\sqrt{2}}{5}$&$g_{\Xi_{c}^{\prime}\Xi_{c}^{\prime}\pi}$  & $\frac{4}{5\sqrt{2}}$&$g_{\Xi_{c}^{\prime}\Xi_{c}\eta}$  & $-\frac{\sqrt{18}}{5}$\\
  \hline

$f_{NN\rho}$ 
&$f_{\Xi_{c}^{\prime}\Xi_{c}^{\prime}\rho}$  & $\frac{1}{2\sqrt{2}} $ &
$f_{\Xi_{c}^{\prime}\Xi_{c}^{\prime}\omega}$& $\frac{1}{2\sqrt{2}} $ &$f_{\Xi_{c}\Xi D_s^{*}}$ & $ -\frac{\sqrt{3}}{2}$
 \\

&$f_{\Xi_{c}^{\prime}\Xi_{c}\rho}$ & $-\frac{\sqrt{3}}{2\sqrt{2}}$ &
$f_{\Xi_{c}^{\prime}\Xi_{c}\omega}$  &$-\frac{\sqrt{3}}{2\sqrt{2}}$ &$f_{\Xi_c\Sigma_c \bar{K}^{*}}$ &$\frac{\sqrt{3}}{2\sqrt{2}}$ \\

&$f_{\Xi_c^{\prime}\Sigma_c \bar{K}^{*}}$&$-\frac{\sqrt{3}}{2\sqrt{2}}$&$f_{\Xi_c\Omega_c K^{*}}$&$-\frac{\sqrt{3}}{2}$&$f_{\Xi_c^{\prime}\Omega_c K^{*}}$ & $\frac{1}{2} $\\

&$f_{\Xi_{c}\Sigma D^{*}}$ & $ \frac{\sqrt{3}}{2\sqrt{2}} $&$f_{\Xi_{c}\Lambda D^{*}}$ & $ \frac{1}{2\sqrt{2}}$&$f_{\Lambda_c ND^*}$&$-\frac{\sqrt{3}}{2}$\\

&$f_{\Xi_{c}^{\prime}\Sigma D^{*}}$ & $ \frac{1}{2\sqrt{2}}$ &$f_{\Xi_{c}^{\prime}\Lambda D^{*}}$ & $ -\frac{\sqrt{3}}{2\sqrt{2}}$ &$f_{\Sigma_c\Sigma_c\rho}$&$\frac{1}{2}$\\
&$f_{\Sigma_c\Sigma_c\omega}$&$\frac{1}{2}$&$f_{\Sigma_c\Lambda_c\rho}$
&$\frac{\sqrt{3}}{2}$&$f_{\Lambda_c\Lambda D_s^*}$&$\frac{1}{\sqrt{2}}$\\
&$f_{\Sigma_cND^*}$&$\frac{1}{2}$ &$f_{\Lambda N \bar{K}^*}$&$-\frac{\sqrt{3}}{2}$&$f_{\Sigma N \bar{K}^*}$&$\frac{1}{2}$\\
&$f_{\Sigma_c\Sigma D_s^*}$ &$-\frac{\sqrt{2}}{2}$&$f_{\Lambda_c\Xi_c^{\prime} \bar{K}^{*}}$ & $\frac{\sqrt{3}}{2\sqrt{2}}$&$f_{\Lambda_c\Lambda_c\omega}$&$-\frac{1}{2}$\\
&$f_{\Xi_{c}^{\prime}\Xi D_s^{*}}$ & $ \frac{1}{2}$ &\\
\hline

$g_{\Delta N\rho}$
&$g_{\Xi_c^{\prime}\Omega_c^{*} K^{*}}$ & $-\frac{1}{\sqrt{6}}$    &$g_{\Xi_c\Omega_c^{*} K^{*}}$ & $-\frac{1}{\sqrt{2}}$
  & $g_{\Xi^{*}\Sigma \bar{K}^{*}}$ & $ -\frac{1}{2\sqrt{3}}$ \\
&$g_{\Xi^{*}\Lambda \bar{K}^{*}}$ & $ -\frac{1}{\sqrt{2}}$&$g_{\Xi^{*}\Xi_{c}^{\prime} D_{s}^{*}}$ & $ \frac{2}{\sqrt{6}}$&$g_{\Sigma_c\Sigma^*_c\rho}$&$\frac{1}{\sqrt{6}}$ \\
&$g_{\Sigma_c\Sigma^*_c\omega}$&$-\frac{1}{\sqrt{6}}$ &$g_{\Sigma_c\Xi_c^* \bar{K}^*}$ & $-\frac{1}{2\sqrt{3}}$&$g_{\Lambda_c\Xi_c^* \bar{K}^*}$ & $\frac{1}{2}$\\
&$g_{\Xi_{c}\Xi_{c}^{*} \rho}$ & $-\frac{1}{2}$ &$g_{\Xi_{c}\Xi_{c}^{*} \omega}$ & $-\frac{1}{2}$&$g_{\Xi_{c}^{\prime}\Xi_{c}^{*} \rho}$ & $-\frac{1}{2\sqrt{3}}$\\
&$g_{\Xi_{c}^{\prime}\Xi_{c}^{*} \pi}$ & $-\frac{1}{2\sqrt{3}}$\\
\hline

$g_{\Delta N\pi}$ 
& $g_{\Xi_c^{\prime} \Omega_c^{*} K}$ & $-\frac{1}{\sqrt{6}}$ 
  &$g_{\Xi_{c}\Xi_{c}^{*} \pi}$ & $-\frac{1}{2}$  &$g_{\Xi^{*}\Xi_{c}^{\prime} D_{s}}$ & $ \frac{2}{\sqrt{6}}$  \\
&$g_{\Xi_c\Omega_c^{*} K}$ & $-\frac{1}{\sqrt{2}}$&$g_{\Sigma_c\Sigma^*_c\pi}$&$\frac{1}{\sqrt{6}}$& $g_{\Sigma_c\Sigma^*_c\eta}$&$\frac{1}{2\sqrt{2}}$ \\
&$g_{\Sigma_c\Xi_c^* \bar{K}}$ & $-\frac{1}{2\sqrt{3}}$&$g_{\Xi_{c}\Xi_{c}^{*} \eta}$ & $-\frac{1}{2}$ &$g_{\Xi_{c}^{\prime}\Xi_{c}^{*} \pi}$ & $-\frac{1}{2\sqrt{3}}$ \\
&$g_{\Xi_{c}^{\prime}\Xi_{c}^{*} \eta}$ & $-\frac{1}{2\sqrt{3}}$\\
\hline

$g_{NN\rho}$ 
&$g_{\Xi_{c}^{\prime}\Xi_{c}^{\prime}\rho}$ & $\sqrt{2}\alpha_{BBV} $ &$g_{\Xi_{c}^{\prime}\Xi_{c}^{\prime}\omega}$ & $\sqrt{2}\alpha_{BBV}$&$g_{\Xi_c^{\prime}\Sigma_c \bar{K}^{*}}$ &$\sqrt{2}{\alpha_{BBV}}$\\
 & $g_{\Sigma_c\Sigma_c\rho}$ & $2\alpha_{BBV}$&$g_{\Sigma_c\Sigma_c\omega}$&$2\alpha_{BBV}$&$g_{\Xi_c^{\prime}\Omega_c K^{*}}$ & $2\alpha_{BBV} $\\
&$g_{\Xi_{c}^{\prime}\Xi_{c}\rho}$ & $\sqrt{\frac{2}{3}}(\alpha_{BBV}-1) $ &
$g_{\Xi_{c}^{\prime}\Xi_{c}\omega}$  & $\sqrt{\frac{2}{3}}(\alpha_{BBV}-1) $ & $g_{\Xi_c\Omega_c K^{*}}$&$\frac{2}{\sqrt{3}}(\alpha_{BBV}-1) $ \\

&$g_{\Xi_c\Sigma_c \bar{K}^{*}}$ &$-\sqrt{\frac{2}{3}}(\alpha_{BBV}-1)$ &$g_{\Xi_{c}^{\prime}\Lambda D^{*}}$ & $ \frac{\sqrt{3}}{\sqrt{2}}(2\alpha_{BBV}-1)$& $g_{\Xi_{c}\Xi D_s^{*}}$ & $ -\frac{1}{\sqrt{3}}(2\alpha_{BBV}+1)$\\

&$g_{\Lambda_c ND^*}$&$\frac{1}{3\sqrt{2}}(2\alpha_{BBV}+1)$&$g_{\Xi_{c}^{\prime}\Xi D_s^{*}}$    & $-(2\alpha_{BBV}-1)$ &$g_{\Xi_{c}\Sigma D^{*}}$ & $ \frac{1}{\sqrt{6}}(2\alpha_{BBV}+1)$\\

&$g_{\Sigma_cND^*}$ & $-(2\alpha_{BBV}-1)$&$g_{\Sigma_c\Sigma D_s^*}$&$\sqrt{2}(2\alpha_{BBV}-1)$&$g_{\Lambda N K^*}$&$-\frac{1}{\sqrt{3}}(2\alpha_{BBV}+1)$\\
&$g_{\Sigma N K^*}$ &$-(2\alpha_{BBV}-1)$&$g_{\Lambda_c\Xi_c^{\prime} \bar{K}^{*}}$ & $-\frac{\sqrt{2}}{\sqrt{3}}(\alpha_{BBV}-1)$&$g_{\Lambda_c\Lambda_c\omega}$&$\frac{2}{3}(5\alpha_{BBV}-2)$\\
&$g_{\Lambda_c \Lambda D_s^*}$&$\frac{\sqrt{2}}{3}(2\alpha_{BBV}+1)$&$g_{\Xi_{c}^{\prime}\Sigma D^{*}}$ & $ -\frac{1}{\sqrt{2}}(2\alpha_{BBV}-1)$&$g_{\Sigma_c\Lambda_c\rho}$&$-\frac{2}{\sqrt{3}}(\alpha_{BBV}-1)$\\

&$g_{\Xi_{c}\Lambda D^{*}}$ & 
$\frac{1}{3\sqrt{2}}(2\alpha_{BBV}+1)$&\\
\hline

\bottomrule[1pt]\bottomrule[1pt]
\end{tabular}
\end{table*}

\end{document}